\documentclass[12pt]{article}

\usepackage{a4}
\usepackage{color}
\usepackage{cite}
\usepackage{pslatex}
\usepackage{txfonts}
\usepackage{axodraw}
\usepackage{epsfig}

\def\arraystretch{1.2}

\makeatletter
\@addtoreset{equation}{section}
\@addtoreset{figure}{section}
\@addtoreset{table}{section}
\makeatother
\renewcommand\theequation{%
  \ifnum \value{section}>0
     \thesection.\arabic{equation}%
  \else
     \arabic{equation}%
  \fi}
\renewcommand\thefigure{%
  \ifnum \value{section}>0
     \thesection.\arabic{figure}%
  \else
     \arabic{figure}%
  \fi}
\renewcommand\thetable{%
  \ifnum \value{section}>0
     \thesection.\arabic{table}%
  \else
     \arabic{table}%
  \fi}

\def\uspinor{u}
\def\vspinor{{\varv}}
\def\GluonPol{{\varepsilon}}
\def\MSbar{\ensuremath{\overline{\mbox{MS}}}\ }
\makeatletter
\def\make@slash#1#2{\setbox\z@\hbox{$#1#2$}%
  \hbox to 0pt{\hss$#1/$\hss\kern-\wd0}\box0}
\def\dsl{\mathpalette\make@slash}
\makeatother

\def\nn{\nonumber}

\def\Vone{{ Version 1\ }}
\def\Vtwo{{ Version 2\ }}

\newcommand{\rT}{{\mathrm{T}}}
\newcommand{\ri}{{\mathrm{i}}}

\def\GeV{\unskip\,\mathrm{GeV}}
\def\MeV{\unskip\,\mathrm{MeV}}
\def\pba{\unskip\,\mathrm{pb}}

\def\refeq#1{\mbox{Eq.~(\ref{#1})\ }}

\def\reffi#1{\mbox{Figure~\ref{#1}}}

\def\refta#1{\mbox{Table~\ref{#1}}}
\def\reftas#1{\mbox{Tables~\ref{#1}}}
\def\refse#1{\mbox{Section~\ref{#1}}}

\def\citere#1{\mbox{Ref.~\cite{#1}}}
\def\citeres#1{\mbox{Refs.~\cite{#1}}}

\def\mathswitchr#1{\relax\ifmmode{\mathrm{#1}}\else$\mathrm{#1}$\fi}
\newcommand{\Pp}{\mathswitchr p}
\newcommand{\Pt}{\mathswitchr t}

\newcommand{\Pg}{\mathswitchr g}
\newcommand{\PH}{\mathswitchr H}

\def\mathswitch#1{\relax\ifmmode#1\else$#1$\fi}
\newcommand{\Mt}{\mathswitch {m_\Pt}}

\newcommand{\gs}{g_{\mathrm{s}}}

\def\ie{i.e.\ }

\newcommand{\NLO}{{\mathrm{NLO}}}

\def\onejet{{\mbox{1-jet}}}
\newcommand{\ptjet}{p_{\mathrm{T,jet}}}
\newcommand{\ptjetcut}{p_{\mathrm{T,jet,cut}}}

\begin{document}
\enlargethispage{2cm}
\thispagestyle{empty}
\def\thefootnote{\fnsymbol{footnote}}
\setcounter{footnote}{1}
\null
\hfill MPP-2008-98\\
\strut\hfill MZ-TH/08-28\\
\strut\hfill TTP-08-44\\ 
\strut\hfill SFB-CPP-08-79\\
\vspace{1.cm}
\begin{center}
{\Large \bf
  Hadronic top-quark pair production \\
  in association with a hard jet at next-to-leading order QCD: \\
  Phenomenological studies  for the Tevatron and the LHC
  \par}
\vspace{1.3cm}
{\large
  {\sc S.\ Dittmaier$^1$, P.\ Uwer$^2$ and S.\ Weinzierl$^3$} } \\[.5cm]
$^1$ {\it Max-Planck-Institut f\"ur Physik
(Werner-Heisenberg-Institut), \\
D-80805 M\"unchen, Germany}
\\[.5cm]
$^2$ {\it Institut f\"ur Theoretische Teilchenphysik,\\
Universit\"at Karlsruhe, D-76128 Karlsruhe, Germany}
\\[.5cm]
$^3$ {\it Institut f\"ur Physik, \\
Universit\"at Mainz, D-55099 Mainz, Germany}
\par \vskip 3cm
\end{center}\par
We report on the calculation of the next-to-leading order QCD corrections
to the production of top--antitop-quark pairs in association with a hard
jet at the Tevatron and at the LHC. 
Results for integrated and differential cross sections are presented.
We find a significant reduction of the scale dependence. In most cases
the corrections are below 20\% indicating that the perturbative
expansion is well under control.
Moreover,
the forward--backward charge asymmetry of the top-quark, which is analyzed
at the Tevatron, is studied at next-to-leading order. 
We find large corrections suggesting that the definition of the
observable has to be refined.

\vfill
\noindent
October 2008 
\null
\setcounter{page}{0}
\clearpage
\def\thefootnote{\arabic{footnote}}
\setcounter{footnote}{0}

\section{Introduction}

The top-quark is by far the heaviest elementary fermion in the 
Standard Model (SM). With a mass of $(172.6\pm 1.4)$ GeV \cite{Heintz:2008ue} 
its mass is about 36 times larger than the mass of the next heaviest
fermion, the bottom quark. 
The large mass has lead to various speculations whether the top-quark 
behaves as a normal quark or whether it plays a special role in
particle physics. 
The electroweak SU(2)$\times$U(1) gauge structure of the SM, 
which is successful in describing a large variety of measurements, 
would require the quark masses to be zero if the symmetry was
un-broken. With the largest mass amongst the quarks it is thus natural 
to assume that the top-quark is most sensitive to the mechanism 
of electroweak symmetry breaking.
In particular, the fact that the top-quark mass is close to the scale of
electroweak symmetry breaking---or equivalently, that the Yukawa coupling to
the Higgs is very close to one---has motivated different scenarios
in which the top-quark drives the electroweak symmetry breaking. 
More details can be found in recent review 
articles~\cite{Bernreuther:2008ju,Han:2008xb}. 

Ignoring the SM as {\it the theory} of particle physics one might 
still wonder whether the top-quark, which is almost as heavy as a gold
atom, behaves as a point-like particle. 
A deviation from the point-like nature would appear as anomalous 
moments yielding differential distributions different from
the point-like case. Anomalous couplings to the gluon are most
naturally probed via the production of an additional jet. An indirect 
measurement through the measurement of the total cross section is
in general more difficult. This is in particular true when the 
interference term with the corresponding Born amplitude gives no
contribution to the total cross section due to discrete symmetries.   

In the context of the SM we have the remarkable fact that the
electroweak top-quark interactions are
completely determined through the aforementioned SU(2)$\times$U(1) 
gauge structure of the SM.
The only free parameter appearing in top-quark
physics is thus the top-quark mass or equivalently the Yukawa coupling to the
Higgs boson. Once this parameter is measured all
remaining properties are predicted. 

An important task for
the ongoing Tevatron collider and the recently started LHC is the precise
measurement of the top-quark properties. The ultimate goal is to
measure the spin and the quantum numbers of the top-quark as precisely
as possible. Any deviation from the SM would signal new physics. 
There is a variety of measurements which are currently done at the
Tevatron and will be continued at the LHC. The total cross section,
which is currently measured at the Tevatron with an accuracy of about
10\% \cite{ICHEP:2008}, is expected to be measured at the LHC with an accuracy
of 5\%. This measurement allows to make precise tests of the production
mechanism. Other important measurements comprise the cross section for
single top-quark production \cite{Abazov:2006gd,Abazov:2008kt,Aaltonen:2008sy,Harris:2002md,Sullivan:2004ie,Kidonakis:2006bu,Kidonakis:2007ej}, 
the W-polarisation in top-quark decay or the
spin correlations of top-quark pairs 
\cite{Affolder:1999mp,Acosta:2004mb,Abulencia:2005xf,Abulencia:2006ei,Abazov:2007ve,Abazov:2006hb,Abazov:2004ym,Bernreuther:2001rq,Bernreuther:2004jv}. 
Of great interest is also the electric charge of the top-quark and
its couplings to the Z-boson and the Higgs boson. 
They can be constrained via the measurements
of the cross sections for $\Pt\bar\Pt \gamma$,
$\Pt\bar\Pt Z$~\cite{Lazopoulos:2008de}, 
and $\Pt\bar\Pt H$~\cite{Beenakker:2001rj,Beenakker:2002nc,%
Dawson:2002tg,Dawson:2003zu} production. 

The production of a top-quark pair together with an additional jet is
a further important reaction. This becomes already clear from the simple
observation that a substantial number of events in the inclusive 
top-quark sample is accompanied by an additional jet. Depending on the
energy of the additional jet the fraction of events with an additional
jet can easily be of the order of 10--30\% or even more. For example
at the LHC we find a cross section of 376~pb for the production of a 
top--antitop-quark pair with an additional jet with a transverse momentum
above 50~GeV. This is almost half of the total top-quark pair cross
section which is 806~pb~\cite{Moch:2008qy} if evaluated in 
next-to-leading order (NLO)\footnote{Both numbers correspond to a top mass of $174\;\mbox{GeV}$. 
The number for top--antitop-quark pair production with an additional jet was obtained using
CTEQ6M as pdf set, the one for the total top-quark pair cross section with CTEQ6.5}.
For a more precise understanding of the topology of top-quark events
it is thus important to have also an improved understanding of 
top-quark pair production together with a jet. As mentioned already
above, this reaction provides a sensitive tool to search for
anomalous top-quark--gluon couplings. The emission of an additional
gluon also leads to a rather interesting property of the cross
section: The differential cross section contains contributions from
the interference of C-odd and C-even parts of the amplitude\cite{Halzen:1987xd,Kuhn:1998kw,Kuhn:1998jr,Bowen:2005ap}, where
C denotes the charge conjugation (for a similar effect in QED see for
example \citeres{Berends:1973fd,Berends:1982dy}). 
While for the total cross section these
contributions cancel when integrating over the (symmetric) phase space
they can lead to a forward--backward charge 
asymmetry of the top-quark which is measured
at the Tevatron \cite{:2007qb,Aaltonen:2008hc}. It should be stressed that no
parity-violating interactions are involved. Note that the
naively defined forward--backward charge asymmetry is zero at the LHC due to
the symmetric initial state. 
A definition that leads to a non-trivial prediction here requires to select
a preferred axis for each event\cite{Antunano:2007da}, but it is not yet
clear whether an asymmetry survives that is significant over all uncertainties.
In inclusive top-quark pair production at the Tevatron 
the charge asymmetry appears first at one loop, because it
results from interferences of C-odd with C-even parts of double-gluon 
exchange between initial and final states. 
The asymmetry for the inclusive sample has been studied in detail in 
\citeres{Halzen:1987xd,Kuhn:1998kw,Kuhn:1998jr,Bowen:2005ap}.
The available predictions for $\Pt\bar\Pt$
production---although of one-loop order---describes 
this asymmetry only at leading-order (LO) accuracy in QCD.
Recently the analysis has been extended to take large threshold
logarithms at the next-to-leading-log (NLL) level into account 
\cite{Almeida:2008ug}. It was found that at least this class of
higher-order contributions do not change the
theoretical prediction dramatically. The main reason is that this type of
corrections affect the asymmetric cross section roughly in the same way as
the symmetric one. In the ratio the corrections thus cancel to a large
extent and lead to 
a stable theoretical prediction. 
In $\Pt\bar\Pt{+}$jet production the asymmetry appears already in 
the tree amplitude.
Thus, the NLO 
calculation described in this article provides a true
NLO prediction for the asymmetry.
The calculation presented in this work is an important tool in the 
experimental analysis of this observable at the Tevatron where 
the asymmetry is measured \cite{:2007qb,Aaltonen:2008hc}.
In a previous letter \cite{Dittmaier:2007wz} we reported that the 
asymmetry receives large corrections.
In this paper we study the situation in more detail for various
values of the lower cut on the transverse momentum of the hard tagging
jet.

As aforementioned it is expected that the total cross section for 
top-quark pair production will be measured at the LHC with an accuracy
of the order of 5\%. Recently it has been shown  
in \citeres{Moch:2008qy,Cacciari:2008zb,Kidonakis:2008mu} that the
accuracy of the currently available NLO predictions 
is only at the level of 12\% 
(at NLO, but further reduced by the inclusion of the threshold logarithms)
and largely dominated by the scale uncertainty. In \citere{Moch:2008qy}
an estimate to the next-to-next-to-leading order (NNLO) cross section 
has been given. The approximation is
based on the assumption that the NNLO corrections will be dominated by
the threshold region as it is the case for the NLO corrections. 
In the threshold region the logarithmic behaviour together with the two-loop
Coulomb singularity is derived from general arguments. 
In addition the 
complete scale dependence at two loops is included in the approximation.   
Using this approximation to the full NNLO result it is shown in 
\citere{Moch:2008qy} that the theoretical uncertainty may decrease to
a few per cent. The remaining scale uncertainty is of the same order as
the uncertainty induced by the parton distribution functions.
Recently some progress towards a complete NNLO calculation
has been made 
\cite{Korner:2005rg,Korner:2008bn,Moch:2007pj,Czakon:2007ej,Czakon:2007wk,%
Czakon:2008zk,Bonciani:2008az,Anastasiou:2008vd,Kniehl:2008fd}. 
The one-loop corrections to $\Pt \bar\Pt + \onejet$ constitute an 
important ingredient to the NNLO calculation of $\Pt \bar \Pt$
production at hadron colliders.
In this context we mention that---besides our NLO calculation
presented here and in \citere{Dittmaier:2007wz}---part of the 
one-loop amplitudes to $\Pg\Pg\to\Pt\bar\Pt\Pg$
have also been evaluated in \citere{Ellis:2008ir}.

Apart from its
significance as signal process it turns out that $\Pt \bar \Pt +
\onejet$ production is also an important background to 
various new physics searches. A prominent example is Higgs production
via vector-boson fusion. This reaction represents an important
discovery channel for a SM Higgs boson 
with a mass of up to several $100\GeV$~\cite{Asai:2004ws,Abdullin:2005yn}.
The major background to this reaction is due to 
$\Pt\bar\Pt+\onejet$ \cite{Alves:2003vp}, again underlining
the need for precise theoretical predictions for this process.

It is well known that predictions at LO in the 
coupling constant of QCD are plagued 
by large uncertainties. In many
cases the LO predictions in QCD give 
only a rough estimate. Only by including NLO corrections
a quantitatively reliable prediction can be obtained. Given that the
conceptual problems in such calculations are solved since quite
some time, one might think that doing the required calculations should
be a straightforward task. Unfortunately it turns out that this is not
the case. The calculation of radiative 
corrections for $2 \to 3$ and  $2 \to 4$
reactions is still non-trivial---not speaking about
reactions with an even higher multiplicity.%
\footnote{More details and references on problems and suggested solutions 
can, for instance, be found in reports like 
\citeres{Buttar:2006zd,Bern:2008ef}.}
The complexity of the corresponding matrix elements
renders computer codes quite lengthy and CPU time consuming. 
The (more or less) automatically generated code
may in addition lead to numerical instabilities.
In particular, the reduction of one-loop tensor integrals to scalar
one-loop integrals is in general difficult to do in a numerically stable
way. 
In that context the calculation of the one-loop corrections
to top-quark pair production with an additional jet is also
interesting as a benchmark process for the development of new
methods. 

In this paper we extend our previous work~\cite{Dittmaier:2007wz} on the
NLO QCD corrections to $\Pt\bar\Pt{+}$jet production at hadron colliders,
where we discussed the scale dependence of the integrated cross sections
at the Tevatron and the LHC and of the top-quark charge asymmetry at
the Tevatron.
We supplement this discussion upon including more numerical results
showing the dependence on the lower cut set on the transverse momentum
of the hard tagging jet and present first results on
differential distributions.
Moreover, we provide numerical results on the virtual one-loop and 
real-emission corrections for
single phase-space points, in order to facilitate future comparisons
to our calculation.

This article is organized as follows:
In \refse{sect:calculation} we briefly describe the calculation
of the NLO corrections. 
Numerical results are presented in \refse{sect:results}. 
In the appendices we provide numerical results on the virtual and real
corrections for individual phase-space points; 
moreover, we collect the tables with the results for the
differential cross sections there.

\section{Outline of the calculation}
\label{sect:calculation}

\subsection{Born approximation}

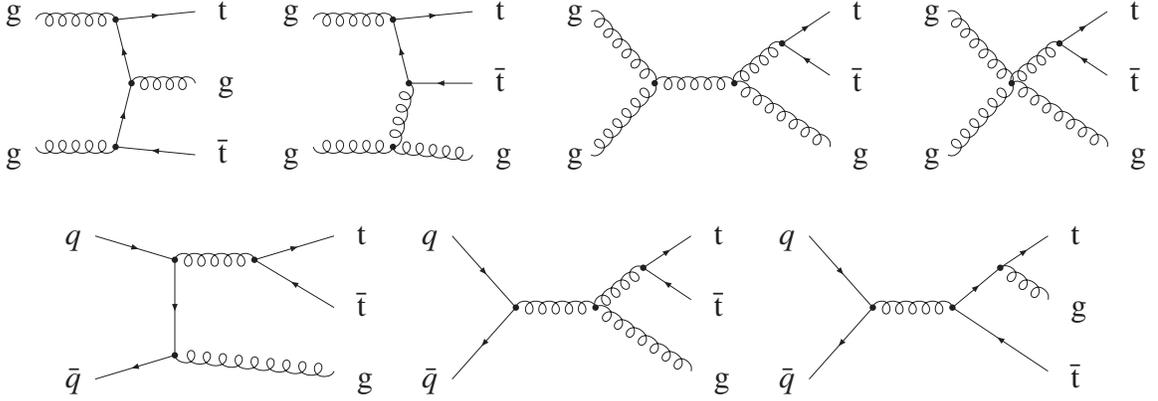
\begin{figure}
\centerline{
{\unitlength .6pt \normalsize
\begin{picture}(120,120)(0,0)
\SetScale{.6}
\Gluon(15, 90)(65,90){4}{5}
\Gluon(65, 10)(15,10){4}{5}
\ArrowLine(65, 10)(75,50)
\ArrowLine(75, 50)(65,90)
\Gluon(75, 50)(115,50){4}{4}
\ArrowLine(115,  5)(65, 10)
\ArrowLine(65,90)(115,95)
\Vertex(65,90){2}
\Vertex(65,10){2}
\Vertex(75,50){2}
\put(-4,90){$\Pg$}
\put(-4, 0){$\Pg$}
\put(130, 90){{$\Pt$}}
\put(130,45){{$\Pg$}}
\put(130, 0){{$\bar{\Pt}$}}
\SetScale{1}
\end{picture}
}
\hspace*{2em}
{\unitlength .6pt \normalsize
\begin{picture}(120,120)(0,0)
\SetScale{.6}
\Gluon(15, 90)(65,90){4}{5}
\Gluon(65, 10)(15,10){4}{5}
\Gluon(75,50)(65, 10){4}{4}
\ArrowLine(75, 50)(65,90)
\ArrowLine(115,50)(75, 50)
\Gluon(115,  5)(65, 10){4}{5}
\ArrowLine(65,90)(115,95)
\Vertex(65,90){2}
\Vertex(65,10){2}
\Vertex(75,50){2}
\put(-4,90){$\Pg$}
\put(-4, 0){$\Pg$}
\put(130, 90){{$\Pt$}}
\put(130,45){{$\bar{\Pt}$}}
\put(130, 0){{$\Pg$}}
\SetScale{1}
\end{picture}
}
\hspace*{2em}
{\unitlength .6pt \normalsize
\begin{picture}(170,120)(0,0)
\SetScale{.6}
\Gluon(15, 95)(55,50){4}{6}
\Gluon(55, 50)(15, 5){4}{6}
\Vertex(55,50){2}
\Gluon(55, 50)(105,50){4}{5}
\Vertex(105,50){2}
\Gluon(105,50)(165, 10){4}{7}
\Gluon(105,50)(135, 75){4}{4}
\ArrowLine(165,55)(135, 75)
\ArrowLine(135,75)(165,95)
\Vertex(135,75){2}
\put(-0,90){$\Pg$}
\put(-0, 0){$\Pg$}
\put(180,90){{$\Pt$}}
\put(180, 0){{$\Pg$}}
\put(180,45){{$\bar{\Pt}$}}
\SetScale{1}
\end{picture}
}
\hspace*{2em}
{\unitlength .6pt \normalsize
\begin{picture}(120,120)(0,0)
\SetScale{.6}
\Gluon(15, 95)(55,50){4}{6}
\Gluon(55, 50)(15, 5){4}{6}
\Vertex(55,50){2}
\Gluon(55,50)(115, 10){4}{7}
\Gluon(55,50)( 85, 75){4}{4}
\ArrowLine(115,55)( 85, 75)
\ArrowLine( 85,75)(115,95)
\Vertex( 85,75){2}
\put(-0,90){$\Pg$}
\put(-0, 0){$\Pg$}
\put(130,90){{$\Pt$}}
\put(130, 0){{$\Pg$}}
\put(130,45){{$\bar{\Pt}$}}
\SetScale{1}
\end{picture}
}
}
\vspace{1em}
\centerline{
{\unitlength .6pt \normalsize
\begin{picture}(170,120)(0,0)
\SetScale{.6}
\ArrowLine(15, 95)(65,80)
\ArrowLine(65, 20)(15, 5)
\Gluon(65, 80)(115,80){4}{5}
\Gluon(165, 10)(65,20){4}{9}
\ArrowLine(65,80)(65,20)
\Vertex(65,80){2}
\Vertex(65,20){2}
\ArrowLine(165,50)(115,80)
\ArrowLine(115,80)(165,95)
\Vertex(115,80){2}
\put(-4,90){$q$}
\put(-4, 0){$\bar q$}
\put(180, 90){{$\Pt$}}
\put(180,45){{$\bar{\Pt}$}}
\put(180, 0){{$\Pg$}}
\SetScale{1}
\end{picture}
}
\hspace*{2em}
{\unitlength .6pt \normalsize
\begin{picture}(170,120)(0,0)
\SetScale{.6}
\ArrowLine(15, 95)(55,50)
\ArrowLine(55, 50)(15, 5)
\Vertex(55,50){2}
\Gluon(55, 50)(105,50){4}{5}
\Vertex(105,50){2}
\Gluon(105,50)(165, 10){4}{7}
\Gluon(105,50)(135, 75){4}{4}
\ArrowLine(165,55)(135, 75)
\ArrowLine(135,75)(165,95)
\Vertex(135,75){2}
\put(-4,90){$q$}
\put(-4, 0){$\bar q$}
\put(180,90){{$\Pt$}}
\put(180, 0){{$\Pg$}}
\put(180,45){{$\bar{\Pt}$}}
\SetScale{1}
\end{picture}
}
\hspace*{2em}
{\unitlength .6pt \normalsize
\begin{picture}(170,120)(0,0)
\SetScale{.6}
\ArrowLine(15, 95)(55,50)
\ArrowLine(55, 50)(15, 5)
\Vertex(55,50){2}
\Gluon(55, 50)(105,50){4}{5}
\Vertex(105,50){2}
\ArrowLine(165, 10)(105,50)
\ArrowLine(105,50)(135, 75)
\Gluon(135, 75)(165,55){4}{3}
\ArrowLine(135,75)(165,95)
\Vertex(135,75){2}
\put(-4,90){$q$}
\put(-4, 0){$\bar q$}
\put(180,90){{$\Pt$}}
\put(180,45){{$\Pg$}}
\put(180, 0){{$\bar{\Pt}$}}
\SetScale{1}
\end{picture}
}
}
\caption{Representative sets of LO diagrams for $\Pg\Pg$ fusion and 
$q\bar q$ annihilation in hadronic $\Pt\bar\Pt{+}$jet production.}
\label{fig:LOgraphs}
\end{figure}

In Born approximation the partonic reactions are 
\begin{equation}
\label{eq:lo_processes}
  \Pg\Pg\to \Pt\bar \Pt \Pg, \quad
  q\bar q \to \Pt\bar \Pt \Pg, \quad
  q\Pg \to \Pt\bar \Pt q, \quad 
  \Pg\bar q \to \Pt\bar \Pt \bar q.
\end{equation}
The last three reactions are related by crossing. 
Therefore, the required generic matrix elements are
\begin{equation}
0 \to \Pt \bar \Pt \Pg \Pg \Pg, \quad 
0 \to \Pt \bar \Pt q \bar q \Pg.
\end{equation}
In the following we generically denote the external momenta and helicities
with $\{p_i\}$ and $\{\lambda_i\}$ and identify the light partons with
$i=1,2,3$. The letter of a specific parton if used as an argument 
denotes the combination of
spin, momentum, and (if relevant) colour of this parton, 
\ie $\Pt=(p_{\Pt},\lambda_{\Pt},i_{\Pt})$ or 
$\Pg_i=(p_i,\lambda_i,a_i)$.
Representative sets of Born diagrams for the $\Pg\Pg$ and $q\bar q$
channels are depicted in \reffi{fig:LOgraphs}.
In total, there are 16 LO diagrams for $0\to\Pt\bar\Pt\Pg\Pg\Pg$ and
5 for $0\to\Pt\bar\Pt q\bar q\Pg$.
The colour decomposition for a tree amplitude corresponding to the process 
$0 \to \Pt \bar \Pt \Pg \Pg \Pg$ is
\begin{eqnarray}
  {\cal A}_{5}^{(0)}(\Pt,\Pg_1,\Pg_2,\Pg_3,\bar \Pt) 
  & = & 
  g_{\mathrm{s}}^{3} 
  \sum\limits_{\sigma\in S_3} 
  \left( T^{a_{\sigma_1}} T^{a_{\sigma_2}} T^{a_{\sigma_3}} 
  \right)_{i_{\Pt} j_{\bar \Pt}}
  A_{5}^{(0)}(\Pt,\Pg_{\sigma_1},\Pg_{\sigma_2},\Pg_{\sigma_3},\bar \Pt),
\end{eqnarray}
where $g_{\mathrm{s}}$ is the strong coupling constant, $S_3$ the
symmetric group and $\sigma=(\sigma_1,\sigma_2,\sigma_3)\in S_3$ a 
permutation. The generators of
the SU(N) gauge group in the fundamental representation are given by $T^a$,
and $a_i$ is the colour index of gluon $\Pg_i$.
The sum extends over all permutations in $S_3$. Physically this
corresponds to all possible colour orderings of the gluons.
The function $A_{5}^{(0)}(\Pt,\Pg_1,\Pg_2,\Pg_3,\bar \Pt)$ is thus the 
colour-ordered subamplitude often also called {\it partial amplitude}.
Due to the colour-ordering, only diagrams with a particular 
ordering contribute to $A_{5}^{(0)}(\Pt,\Pg_1,\Pg_2,\Pg_3,\bar \Pt)$.
The partial amplitudes contain the kinematic information
and are  individually gauge invariant.
The corresponding colour decomposition for the process 
$0 \to \Pt \bar \Pt q \bar q \Pg$ reads
\def\ArgList{{\Pt,\bar \Pt, q, \bar q,  g, }}
\begin{eqnarray}
  \label{eq:qqttg}
{\cal A}_{5}^{(0)}(\Pt,\bar \Pt, q, \bar q, g) & = & 
 g_{\mathrm{s}}^3
 \left[
 \frac{1}{2} \delta_{i_{\Pt}j_{\bar q}} T^a_{i_{q} j_{\bar \Pt}} 
   A_{5,1}^{(0)}(\ArgList)
 +
 \frac{1}{2} \delta_{i_{q}j_{\bar \Pt}} T^a_{i_{\Pt} j_{\bar q}} 
   A_{5,2}^{(0)}(\ArgList)
 \right. \nonumber \\
 & & \left.
 -
 \frac{1}{2N} \delta_{i_{\Pt}j_{\bar \Pt}} T^a_{i_{q} j_{\bar q}} 
   A_{5,3}^{(0)}(\ArgList)
 -
 \frac{1}{2N} \delta_{i_{q}j_{\bar q}} T^a_{i_{\Pt} j_{\bar \Pt}} 
   A_{5,4}^{(0)}(\ArgList)
 \right].  
\end{eqnarray}
We note that the amplitudes $A_{5,i}^{(0)}$ ($i=1,2,3,4$) are linearly
dependent. The relation 
\begin{displaymath}
  0=A_{5,1}^{(0)}+A_{5,2}^{(0)}-A_{5,3}^{(0)}-A_{5,4}^{(0)}
\end{displaymath}
can be used to express for example $A_{5,4}^{(0)}$ in terms of
$A_{5,i}^{(0)}$ with $i=1,2,3$. This is particularly useful for the evaluation
of the squared amplitude.
Compact analytic results for the LO amplitudes are given in 
\citere{Bernreuther:2004jv} where the amplitudes have been used in the
calculation of the NLO corrections for top-quark pair production. 
In addition we also performed several
independent calculations, including one with 
{\sl Madgraph}~\cite{Stelzer:1994ta}, and found complete numerical 
agreement among all those calculations.
  
\subsection{Virtual corrections}

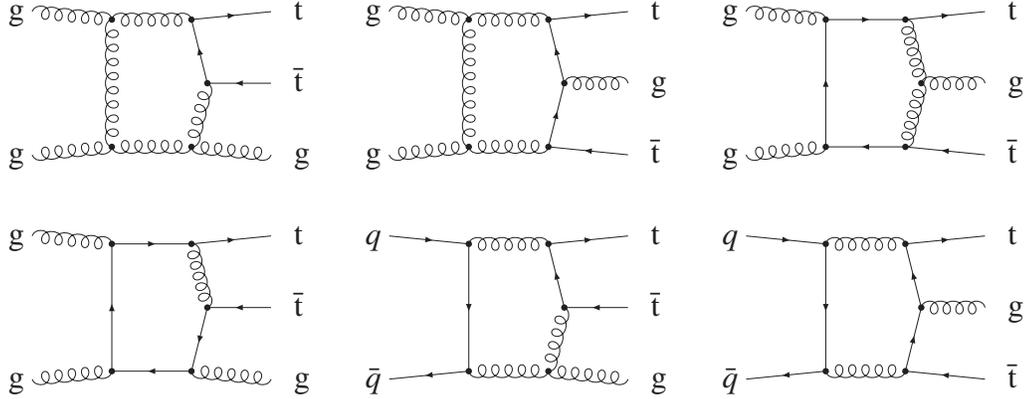
\begin{figure}
\centerline{
{\unitlength .6pt \normalsize
\begin{picture}(170,120)(0,0)
\SetScale{.6}
\Gluon(15, 95)(65,90){4}{5}
\Gluon(65, 10)(15, 5){4}{5}
\Gluon(65, 90)(115,90){4}{5}
\Gluon(115, 10)(65,10){4}{5}
\Gluon(65,10)(65,90){4}{8}
\Vertex(65,90){2}
\Vertex(65,10){2}
\Gluon(125,50)(115, 10){4}{4}
\ArrowLine(125, 50)(115,90)
\ArrowLine(165,50)(125, 50)
\Gluon(165,  5)(115, 10){4}{5}
\ArrowLine(115,90)(165,95)
\Vertex(115,90){2}
\Vertex(115,10){2}
\Vertex(125,50){2}
\put(-0,90){$\Pg$}
\put(-0, 0){$\Pg$}
\put(180, 90){{$\Pt$}}
\put(180,45){{$\bar{\Pt}$}}
\put(180, 0){{$\Pg$}}
\SetScale{1}
\end{picture}
}
\hspace*{2em}
{\unitlength .6pt \normalsize
\begin{picture}(170,120)(0,0)
\SetScale{.6}
\Gluon(15, 95)(65,90){4}{5}
\Gluon(65, 10)(15, 5){4}{5}
\Gluon(65, 90)(115,90){4}{5}
\Gluon(115, 10)(65,10){4}{5}
\Gluon(65,10)(65,90){4}{8}
\Vertex(65,90){2}
\Vertex(65,10){2}
\ArrowLine(115, 10)(125,50)
\ArrowLine(125, 50)(115,90)
\Gluon(125, 50)(165,50){4}{4}
\ArrowLine(165,  5)(115, 10)
\ArrowLine(115,90)(165,95)
\Vertex(115,90){2}
\Vertex(115,10){2}
\Vertex(125,50){2}
\put(-0,90){$\Pg$}
\put(-0, 0){$\Pg$}
\put(180, 90){{$\Pt$}}
\put(180,45){{$\Pg$}}
\put(180, 0){{$\bar{\Pt}$}}
\SetScale{1}
\end{picture}
}
\hspace*{2em}
{\unitlength .6pt \normalsize
\begin{picture}(170,120)(0,0)
\SetScale{.6}
\Gluon(15, 95)(65,90){4}{5}
\Gluon(65, 10)(15, 5){4}{5}
\ArrowLine(65, 90)(115,90)
\ArrowLine(115, 10)(65,10)
\ArrowLine(65,10)(65,90)
\Vertex(65,90){2}
\Vertex(65,10){2}
\Gluon(125,50)(115, 10){4}{5}
\Gluon(115,90)(125, 50){4}{5}
\Gluon(125, 50)(165,50){4}{4}
\ArrowLine(165,  5)(115, 10)
\ArrowLine(115,90)(165,95)
\Vertex(115,90){2}
\Vertex(115,10){2}
\Vertex(125,50){2}
\put(-0,90){$\Pg$}
\put(-0, 0){$\Pg$}
\put(180, 90){{$\Pt$}}
\put(180,45){{$\Pg$}}
\put(180, 0){{$\bar{\Pt}$}}
\SetScale{1}
\end{picture}
}
}
\vspace{1em}
\centerline{
{\unitlength .6pt \normalsize
\begin{picture}(170,120)(0,0)
\SetScale{.6}
\Gluon(15, 95)(65,90){4}{5}
\Gluon(65, 10)(15, 5){4}{5}
\ArrowLine(65, 90)(115,90)
\ArrowLine(115, 10)(65,10)
\ArrowLine(65,10)(65,90)
\Vertex(65,90){2}
\Vertex(65,10){2}
\ArrowLine(125,50)(115, 10)
\Gluon(115,90)(125, 50){4}{5}
\ArrowLine(165,50)(125, 50)
\Gluon(165,  5)(115, 10){4}{5}
\ArrowLine(115,90)(165,95)
\Vertex(115,90){2}
\Vertex(115,10){2}
\Vertex(125,50){2}
\put(-0,90){$\Pg$}
\put(-0, 0){$\Pg$}
\put(180, 90){{$\Pt$}}
\put(180,45){{$\bar{\Pt}$}}
\put(180,0){{$\Pg$}}
\SetScale{1}
\end{picture}
}
\hspace*{2em}
{\unitlength .6pt \normalsize
\begin{picture}(170,120)(0,0)
\SetScale{.6}
\ArrowLine(15, 95)(65,90)
\ArrowLine(65, 10)(15, 5)
\Gluon(65, 90)(115,90){4}{5}
\Gluon(115, 10)(65,10){4}{5}
\ArrowLine(65,90)(65,10)
\Vertex(65,90){2}
\Vertex(65,10){2}
\Gluon(125,50)(115, 10){4}{4}
\ArrowLine(125, 50)(115,90)
\ArrowLine(165,50)(125, 50)
\Gluon(165,  5)(115, 10){4}{5}
\ArrowLine(115,90)(165,95)
\Vertex(115,90){2}
\Vertex(115,10){2}
\Vertex(125,50){2}
\put(-0,90){$q$}
\put(-0, 0){$\bar q$}
\put(180, 90){{$\Pt$}}
\put(180,45){{$\bar{\Pt}$}}
\put(180, 0){{$\Pg$}}
\SetScale{1}
\end{picture}
}
\hspace*{2em}
{\unitlength .6pt \normalsize
\begin{picture}(170,120)(0,0)
\SetScale{.6}
\ArrowLine(15, 95)(65,90)
\ArrowLine(65, 10)(15, 5)
\Gluon(65, 90)(115,90){4}{5}
\Gluon(115, 10)(65,10){4}{5}
\ArrowLine(65,90)(65,10)
\Vertex(65,90){2}
\Vertex(65,10){2}
\ArrowLine(115, 10)(125,50)
\ArrowLine(125, 50)(115,90)
\Gluon(125, 50)(165,50){4}{4}
\ArrowLine(165,  5)(115, 10)
\ArrowLine(115,90)(165,95)
\Vertex(115,90){2}
\Vertex(115,10){2}
\Vertex(125,50){2}
\put(-0,90){$q$}
\put(-0, 0){$\bar q$}
\put(180, 90){{$\Pt$}}
\put(180,45){{$\Pg$}}
\put(180, 0){{$\bar{\Pt}$}}
\SetScale{1}
\end{picture}
}
}
\caption{Representative sets of pentagon diagrams for 
$\Pg\Pg$ fusion and $q\bar q$ annihilation in hadronic
$\Pt\bar\Pt{+}$jet production at NLO QCD.}
\label{fig:pentagons}
\end{figure}

The virtual corrections consist of the one-loop corrections to the
LO reactions. One can classify the
corrections into self-energy, vertex, box-type,
and pentagon-type corrections where all the external legs
are directly connected  to the loop thus forming a pentagon. 
The latter are the most complicated
ones due to their complexity and the involved tensor integrals. 
Typical examples of the pentagon graphs are shown in \reffi{fig:pentagons}.
Specifically, there are 24 pentagons 
for $0\to\Pt\bar\Pt\Pg\Pg\Pg$ and
8 for $0\to\Pt\bar\Pt q\bar q\Pg$. The total number of diagrams is
354 for the $0\to\Pt\bar\Pt\Pg\Pg\Pg$ case and 94
for the $0\to\Pt\bar\Pt q\bar q\Pg$ case.
The challenging step in this context is the numerically fast and stable
reduction of the tensor integrals to scalar one-loop integrals. 

Before describing the details we briefly outline the general
setup. Owing to the involved kinematics the individual Feynman
diagrams lead to large expressions which are cumbersome to 
evaluate.
To be able to handle the large
expressions and to ensure a fast numerical evaluation at the end we
used a decomposition of the amplitude according to the spin and colour 
structure. Schematically the decomposition of the one-loop 
amplitude ${\cal A}^{(1)}_5$ reads
\begin{equation}
  {\cal A}^{(1)}_5 = 
  \sum_{c,s} {\cal C}_c \times {\cal S}_s(\{p_i\},\{\lambda_i\}) 
  \times f^{(1)}_{cs}(\{p_i\cdot p_j\}),
\label{eq:Avirt}
\end{equation}
where ${\cal C}_c$ denote the colour structures, 
${\cal S}_s$ are the spin structures
(elsewhere called ``standard matrix elements''), and
the functions $f^{(1)}_{cs}$ are scalar functions that depend 
only on the scalar
products of the external momenta $p_i$.
In detail, for $0\to\Pt\bar\Pt\Pg\Pg\Pg$ there are
10 independent colour structures
${\cal C}_c$ of the form 
\begin{displaymath}
 \left(
  T^{a_{\sigma_1}} T^{a_{\sigma_2}} T^{a_{\sigma_3}}
\right)_{i_{\Pt} j_{\bar \Pt}},
\qquad
 \left( T^{a_{\sigma_1}} \right)_{i_{\Pt} j_{\bar \Pt}} 
\delta^{a_{\sigma_2}a_{\sigma_3}},
\qquad
\ri f^{a_1a_2a_3 }
\delta_{i_{\Pt} j_{\bar \Pt}},
\end{displaymath}
where the structure constant of the SU(3) gauge group $f^{abc}$ is
defined
in the usual way through
\begin{equation}
[T^a,T^b] = \ri f^{abc} T^c.  
\end{equation}
We note that the counting refers to the group SU(3); for SU(N) there are
11 independent structures. For N=3 the generic structure 
$\mbox{Tr}[T^{a_1}T^{a_2}T^{a_3}]\delta_{i_{\Pt} j_{\bar \Pt}}$ of the
SU(N) case can be further reduced, because   
$d^{a_1a_2a_3 }\delta_{i_{\Pt} j_{\bar \Pt}}$ appearing through
$\mbox{Tr}[T^{a_1}T^{a_2}T^{a_3}]= {1\over 4} (d^{a_1a_2a_3}+if^{a_1a_2a_3})$
is expressible in terms of the other 10 structures  within SU(3) 
representations.
Allowing only the use of relations which are compatible with
conventional dimensional regularisation when simplifying the Lorentz 
structure of the amplitude, we find a few hundred spin structures, such as 
\begin{displaymath}
  [\bar \vspinor_{\bar \Pt} \uspinor_\Pt] \, 
  (\GluonPol_1\cdot\GluonPol_2) \, (\GluonPol_3\cdot p_2),
  \qquad
  [\bar \vspinor_{\bar \Pt} \dsl{\GluonPol}_3 \uspinor_\Pt] \, 
  (\GluonPol_1\cdot\GluonPol_2), \qquad
  [\bar \vspinor_{\bar \Pt} \dsl{\GluonPol}_2 \dsl{\GluonPol}_3
  \uspinor_\Pt] \, 
  (\GluonPol_1\cdot p_3),
\qquad
\mbox{etc.},
\end{displaymath}
where an obvious notation for the Dirac spinors $\bar \vspinor_{\bar \Pt}$,
$\uspinor_t$ and gluon polarization vectors $\GluonPol_i$ is used. 
Restricting to  four dimensions and using explicitly four-dimensional 
helicity techniques the number of independent structures may be further reduced.
Each gluon polarization vector can be ``gauged'' to be orthogonal
to an arbitrary light-like reference vector, a fact that reduces the algebraic
expressions considerably. A useful choice is, e.g., given by the cyclic
set of conditions
\begin{equation}
  p_2\cdot\GluonPol_1 
  = p_3\cdot\GluonPol_2 
  = p_1\cdot\GluonPol_3 = 0,
\end{equation}
which supplements the transversality relations $p_i\cdot\GluonPol_i=0$.
However, more than 100 spin structures still remain 
(in $D$ dimensions) in spite of this
simplification.
For $0\to\Pt\bar\Pt q\bar q\Pg$ there are only 4 colour structures
(as in LO),
\begin{equation}
 \delta_{i_{\Pt}j_{\bar q}} T^a_{i_{q} j_{\bar \Pt}}, \quad 
 \delta_{i_{q}j_{\bar \Pt}} T^a_{i_{\Pt} j_{\bar q}}, \quad 
 \delta_{i_{\Pt}j_{\bar \Pt}} T^a_{i_{q} j_{\bar q}}, \quad 
 \delta_{i_{q}j_{\bar q}} T^a_{i_{\Pt} j_{\bar \Pt}},
\end{equation}
but also more than 100 spin structures, such as
\begin{displaymath}
  [\bar \vspinor_{\bar \Pt} \uspinor_\Pt] \, 
  [\bar \vspinor_{\bar q} \dsl{p}_{\Pt} \uspinor_q] \, 
  (\GluonPol\cdot p_{\Pt}),
  \qquad
  [\bar \vspinor_{\bar \Pt} \uspinor_\Pt] \, 
  [\bar \vspinor_{\bar q} \dsl{\GluonPol} \uspinor_q],
  \qquad
  [\bar \vspinor_{\bar \Pt} \gamma^\mu \uspinor_\Pt] \, 
  [\bar \vspinor_{\bar q} \gamma_\mu \uspinor_q]
  (\GluonPol\cdot p_{\Pt}),
  \qquad
  \mbox{etc.} \, .
\end{displaymath}

We stress that the algebraic reduction of each Feynman diagram 
to the standard form shown in
\refeq{eq:Avirt} proceeds in $D=4-2\epsilon$ space--time dimensions, i.e.\
only Dirac equations, transversality and gauge conditions of 
polarization vectors, and momentum conservation are used. We have
followed two different strategies for the evaluation of the decomposition
shown in \refeq{eq:Avirt}. In one implementation all Feynman diagrams
were first combined and then projected onto the different structures. The
idea behind this is to form gauge independent quantities where some 
cancellations may happen. In the second approach the decomposition is
applied to each individual Feynman diagram. In the latter
the numerical evaluation of the functions $f_{cs}$ is most efficiently
done if the colour structures completely factorizes. For diagrams
without 4-gluon vertices this is trivially the case. Diagrams where
the number $n_4$ of 4-gluon vertices is greater than 0
are decomposed into $3^{n_4}$ terms, each
with its own colour structure. 
Denoting such one-loop (sub)graphs generically $\Gamma$,
their contributions $f^{(\Gamma)}_{cs}$ to the functions $f_{cs}$ are written as
\begin{equation}
f^{(\Gamma)}_{cs} = c^{(\Gamma)}_c f^{(\Gamma)}_s
\end{equation}
with constants $c^{(\Gamma)}_c$, i.e.\ the total colour structure of $\Gamma$
is $\sum_c {\cal C}_c c^{(\Gamma)}_c$.
The colour-stripped functions $f^{(\Gamma)}_s$, which contain the time-consuming
loop functions are the same for all colour channels.
Writing the LO amplitude as
\begin{equation}
{\cal A}^{(0)}_5 = 
  \sum_c  {\cal C}_c A^{(0)}_{5,c} (\{p_i\},\{\lambda_i\}),
\end{equation}
the contribution of $\Gamma$ to the one-loop-corrected spin- and colour-averaged
squared amplitude is evaluated as follows,
\begin{eqnarray}
\lefteqn{
\sum_{\mbox{\scriptsize colour, spin}} 2\mathrm{Re}\left\{
  {{\cal A}_5^{(0)}}^\ast {\cal A}^{(\Gamma)}_5 \right\} }
\nn\\
&=&
2\mathrm{Re}\left\{
  \sum_{c,c'} C_{cc'} c^{(\Gamma)}_{c'}
  \sum_s f^{(\Gamma)}_s(\{p_i\cdot p_j\}) \, \sum_{\{\lambda_i\}}
  A^{(0)}_{5,c} (\{p_i\},\{\lambda_i\})^* \, {\cal S}_s(\{p_i\},\{\lambda_i\}) 
\right\}.
\end{eqnarray}
In detail, the colour correlation matrix
\begin{equation}
  C_{cc'} = \sum_{\mbox{\scriptsize colour}} {\cal C}^\dagger_c {\cal C}_{c'}
\end{equation}
is calculated only once and for all for the whole process,
and the interferences
${A^{(0)}_{5,c}}^\ast {\cal S}_s$ of the LO amplitude and the different 
spin structures ${\cal S}_s$ are only calculated once per phase-space point
for every spin state.
The approach where first all the diagrams are combined and then the
projection is done yields similar formulae. 
The final formula then reads:
\begin{eqnarray}
  \lefteqn{
    \sum_{\mbox{\scriptsize colour, spin}} 2\mathrm{Re}\left\{
      ({\cal A}_5^{(0)})^\dagger {\cal A}_5 \right\} }
\nn\\
&=&
2\mathrm{Re}\left\{\sum_{\{\lambda_i\}}
  \sum_{c,c'}\sum_{s,s'} C_{cc'} 
   f_{cs}(\{p_i\cdot p_j\}) f^{(0)\ast}_{c's'}(\{p_i\cdot p_j\})\, 
   {\cal S}_s(\{p_i\},\{\lambda_i\}) {\cal S}^*_{s'}(\{p_i\},\{\lambda_i\}) 
\right\},
\end{eqnarray}
where $f^{(0)}_{cs}$ are the scalar functions appearing in
the decomposition of the Born amplitude.
To ensure the correctness of our results the two slightly different
approaches were implemented in two complete independent computer codes.
We note at this point that no significant difference between the two
approaches concerning speed and numerical stability  was
observed. We also note that we used as far as possible
different methods and also different tools to obtain the various
ingredients discussed above. In the following we give some details of
the techniques employed in the two implementations.

{\it Version 1} of the virtual corrections is essentially obtained
following the method described in \citere{Beenakker:2002nc}, where
$\Pt\bar\Pt\PH$ production at hadron colliders was considered.
Feynman diagrams and amplitudes have been generated with the
{\sl FeynArts} package \cite{Kublbeck:1990xc,Hahn:2000kx}
and further processed with in-house {\sl Mathematica} routines,
which automatically create an output in {\sl Fortran}.
The infrared (IR), i.e.\ soft and collinear, 
singularities---which are treated in dimensional
regularisation in both calculations---are analytically separated
from the finite remainder in terms of triangle subdiagrams,
as described in \citeres{Beenakker:2002nc,Dittmaier:2003bc}.
This separation, in particular, allows for a transparent evaluation
of so-called rational terms that originate from $D$-dependent terms
multiplying IR divergences, which appear as single or double poles in $(D-4)$.
As generally shown in \citere{Bredenstein:2008zb}, after properly
separating IR from ultraviolet (UV) divergences such rational terms 
originating from IR
divergences completely cancel; this general result is confirmed in our 
explicit calculation.
The tensor integrals appearing in the pentagon diagrams
are directly reduced to box
integrals following \citere{Denner:2002ii}. 
(Similar methods have been proposed in \citere{Binoth:2005ff}.) 
This method does not
introduce inverse Gram determinants in this step, thereby avoiding
notorious numerical instabilities in regions where these determinants
become small. Box and lower-point integrals are reduced
\`a la Passarino--Veltman \cite{Passarino:1978jh} to scalar integrals,
which are either calculated analytically or using the results of
\citeres{'tHooft:1978xw,Beenakker:1988jr,Denner:1991qq}.
Sufficient numerical stability is already achieved in this
way. Nevertheless the integral evaluation is currently further refined
by employing the more sophisticated methods described in
\citere{Denner:2005nn} in order to numerically stabilize the tensor
integrals in exceptional phase-space regions.

{\it Version 2} of the evaluation of loop diagrams starts
with the generation of diagrams and amplitudes via {\sl QGRAF}
\cite{Nogueira:1991ex},
which are then further manipulated with {\sl Form}
\cite{Vermaseren:2000nd} and {\sl Maple} and eventually
automatically translated into {\sl C++} code.
The reduction of the the 5-point tensor integrals to scalar
integrals is performed with an extension of the method described in
\citere{Giele:2004iy}. In this procedure also
inverse Gram determinants of four four-momenta are avoided.
The lower-point tensor integrals are reduced
using an independent implementation of the Passarino--Veltman procedure.
The IR-finite scalar integrals are
evaluated using the {\sl FF} package
\cite{vanOldenborgh:1990wn,vanOldenborgh:1991yc}.
Although the entire procedure is sufficiently stable, further
numerical stabilization of the tensor reduction is planned
following the expansion techniques suggested in \citere{Giele:2004ub}
for exceptional phase-space regions.

As stated above we used dimensional regularisation to regularise 
UV as well as soft and collinear divergences. 
We renormalised the coupling in a mixed scheme where the light
flavours are treated according to the modified minimal subtraction
$\MSbar$, while the top-quark loop of the gluon self-energy
is subtracted at zero momentum. The 
top-quark mass is renormalised in the on-shell scheme. More
specifially we used the renormalisation constants as given for example
in \citere{Beenakker:2002nc}. In these formulae the divergences of UV
and IR origin are separated. They allow us to check UV and IR
finiteness separately.

\subsection{Real corrections}
The generic matrix elements for the real corrections are given by
\begin{equation}
0 \to \Pt \bar \Pt g g g g, 
\quad
0 \to \Pt \bar \Pt q \bar q g g,
\quad
0 \to \Pt \bar \Pt q \bar q q' \bar q',
\quad
0 \to \Pt \bar \Pt q \bar q q \bar q
\end{equation}
with $q\ne q'$.
The various partonic processes are obtained from these matrix elements 
by all possible crossings of light particles into the initial state.
While the crossing symmetry is extremely helpful in constructing the required
amplitudes it should be kept in mind that the large number of possible
channels obtained from the different crossings lead to a significant 
increase in the computational complexity, given that every channel 
has to be integrated over the phase space. 
The amplitude for the process $0 \to \Pt \bar \Pt q \bar q q \bar q$
with identical quarks $q$
can be obtained from the amplitude of the process 
$0 \to \Pt \bar \Pt q \bar q q' \bar q'$
with non-identical quarks $q$ and $q'$:
\begin{eqnarray}
{\cal A}_{6}^{(0)}(\Pt,\bar \Pt, q, \bar q, q, \bar q) 
 & = & 
{\cal A}_{6}^{(0)}(\Pt,\bar \Pt, q, \bar q, q', \bar q') 
-
{\cal A}_{6}^{(0)}(\Pt,\bar \Pt, q, \bar q', q', \bar q).
\end{eqnarray}
The colour decomposition of a tree amplitude corresponding to the process 
$0 \to \Pt \bar \Pt \Pg \Pg \Pg \Pg$ is
\begin{equation}
  {\cal A}_{6}^{(0)}(\Pt,\Pg_1,\Pg_2,\Pg_3,\Pg_4,\bar \Pt) = 
 g_{\mathrm{s}}^4 \sum\limits_{\sigma\in S_4} 
 \left( T^{a_{\sigma_1}} T^{a_{\sigma_2}} T^{a_{\sigma_3}} T^{a_{\sigma_4}} \right)_{i_{\Pt} j_{\bar \Pt}}
A_{6}^{(0)}(\Pt,\Pg_{\sigma_1},\Pg_{\sigma_2},\Pg_{\sigma_3},\Pg_{\sigma_4},\bar \Pt),
\end{equation}
where the sum is over all permutations 
$\sigma=(\sigma_1,\sigma_2,\sigma_3,\sigma_4)$ of the symmetric group $S_4$. 
The colour decomposition for the process 
$0 \to \Pt \bar \Pt q \bar q g \Pg$ reads
\def\ArgList{{\Pt,\bar \Pt,q,\bar q, \Pg_{\sigma_1}, \Pg_{\sigma_2} }}
\begin{eqnarray}
\lefteqn{
  {\cal A}_{6}^{(0)}(\Pt,\bar \Pt, q, \bar q, \Pg_1, \Pg_2) = 
  \frac{g_{\mathrm{s}}^{4}}{2} \sum\limits_{(\sigma_1\sigma_2)\in S_2}
  \biggl[
    \delta_{i_{\Pt}j_{\bar q}} \left( T^{a_{\sigma_1}} T^{a_{\sigma_2}} \right)_{i_{q} j_{\bar \Pt}} 
    A_{6,1}^{(0)}(\ArgList)
  } & & 
\nonumber \\
& &
  +
  T^{a_{\sigma_1}} _{i_{\Pt}j_{\bar q}} T^{a_{\sigma_2}}_{i_{q} j_{\bar \Pt}} 
  A_{6,2}^{(0)}(\ArgList)
  +
  \left( T^{a_{\sigma_1}} T^{a_{\sigma_2}} \right)_{i_{\Pt}j_{\bar q}} \delta_{i_{q} j_{\bar \Pt}} 
  A_{6,3}^{(0)}(\ArgList)
\nonumber \\
& & 
  -
  \frac{1}{N} \delta_{i_{\Pt}j_{\bar \Pt}} \left( T^{a_{\sigma_1}} T^{a_{\sigma_2}} \right)_{i_{q} j_{\bar q}} 
  A_{6,4}^{(0)}(\ArgList
  -
  \frac{1}{N} T^{a_{\sigma_1}} _{i_{\Pt}j_{\bar \Pt}} T^{a_{\sigma_2}}_{i_{q} j_{\bar q}} 
  A_{6,5}^{(0)}(\ArgList)
\nonumber \\
& & 
  -
  \frac{1}{N} \left( T^{a_{\sigma_1}} T^{a_{\sigma_2}} \right)_{i_{\Pt}j_{\bar \Pt}} \delta_{i_{q} j_{\bar q}} 
  A_{6,6}^{(0)}(\ArgList)
\biggr],
\end{eqnarray}
where again the sum is over all permutations of the gluon legs. 
Finally, the colour decomposition of the process $0 \to \Pt \bar \Pt q
\bar q q' \bar q'$ is
\begin{eqnarray}
\lefteqn{
{\cal A}_{6}^{(0)}(\Pt,\bar \Pt, q, \bar q, q', \bar q')
 =
 \frac{g_{\mathrm{s}}^{4}}{4}
 \biggl[
        \delta_{i_{\Pt}j_{\bar q}} \delta_{i_{q} j_{\bar q'}}
\delta_{i_{q'} j_{\bar \Pt}}
          A_{6,1}^{(0)}(\Pt,\bar \Pt, q, \bar q, q', \bar q')
        +
        \delta_{i_{\Pt}j_{\bar q'}} \delta_{i_{q'} j_{\bar q}}
\delta_{i_{q} j_{\bar \Pt}}
          A_{6,1}^{(0)}(\Pt,\bar \Pt, q', \bar q', q, \bar q)
} & &
 \nonumber \\
 & &
      - \frac{1}{N}
        \delta_{i_{\Pt}j_{\bar q}} \delta_{i_{q} j_{\bar \Pt}}
\delta_{i_{q'} j_{\bar q'}}
          A_{6,2}^{(0)}(\Pt,\bar \Pt, q, \bar q, q', \bar q')
      - \frac{1}{N}
        \delta_{i_{\Pt}j_{\bar q'}} \delta_{i_{q'} j_{\bar \Pt}}
\delta_{i_{q} j_{\bar q}}
          A_{6,2}^{(0)}(\Pt,\bar \Pt, q', \bar q', q, \bar q)
\hspace*{25mm}
 \nonumber \\
 & &
      - \frac{1}{N}
        \delta_{i_{q}j_{\bar q'}} \delta_{i_{q'} j_{q}} \delta_{i_{\Pt}
j_{\bar \Pt}}
          A_{6,3}^{(0)}(\Pt,\bar \Pt, q, \bar q, q', \bar q')
      + \frac{1}{N^2}
        \delta_{i_{\Pt}j_{\bar \Pt}} \delta_{i_{q} j_{\bar q}}
\delta_{i_{q'} j_{\bar q'}}
          A_{6,4}^{(0)}(\Pt,\bar \Pt, q, \bar q, q', \bar q')
 \biggr].
\end{eqnarray}

To extract the  IR singularities and for their
combination with the virtual corrections we employ the dipole
subtraction formalism \cite{Catani:1996vz,Phaf:2001gc,Catani:2002hc}.
Specifically, the formulation~\cite{Catani:2002hc} for massive quarks
is used.
At NLO schematically one has the following contributions:
\begin{eqnarray}
\langle O \rangle^{\NLO} & = & 
 \int\limits_{n+1} O_{n+1} d\sigma^{\mathrm{R}} + \int\limits_n O_n d\sigma^{\mathrm{V}} 
 + \int\limits_n O_n d\sigma^{\mathrm{C}}.
\end{eqnarray}
Here
$d\sigma^{\mathrm{R}}$ denotes the real emission contribution,
whose matrix elements are given by the square of the Born amplitudes 
with $6$ partons $| {\cal A}^{(0)}_{6} |^2$,
$d\sigma^{\mathrm{V}}$ is the virtual contribution, whose matrix 
elements are given by the interference term
of the one-loop amplitudes ${\cal A}^{(1)}_{5}$ with $5$ partons with 
the corresponding Born amplitude ${\cal A}^{(0)}_{5}$, and
$d\sigma^{\mathrm{C}}$ denotes a collinear subtraction term, which originates
from the factorisation of the initial-state collinear singularities.
The function $O_n$ defined on the $n$-particle
phase space stands for any prescription ($\theta$-functions for
phase-space cuts, $\delta$-functions for distributions)  
defining an IR-safe observable.
Taken separately, the individual contributions are IR
divergent, and only their sum is finite.
In order to render the individual contributions finite, 
so that the phase-space integrations can be performed by Monte Carlo 
methods, one adds and subtracts a suitably chosen ``counterterm''
$d\sigma^{\mathrm{A}}$:
\begin{eqnarray}
\langle O \rangle^{\NLO} & = & 
 \int\limits_{n+1} \left( O_{n+1} d\sigma^{\mathrm{R}} - O_n d\sigma^{\mathrm{A}} \right)
 + \int\limits_n \left( O_n d\sigma^{\mathrm{V}} + O_n d\sigma^{\mathrm{C}} + O_n \int\limits_1 d\sigma^{\mathrm{A}} \right).
\end{eqnarray}
The matrix element corresponding to the approximation term $d\sigma^{\mathrm{A}}$ is given as a sum over 
dipoles:
\begin{eqnarray}
d\sigma^{\mathrm{A}} & \propto & 
\sum\limits_{\mathrm{pairs}\; i,j} \; \sum\limits_{k \neq i,j} {\cal D}_{ij,k}.
\end{eqnarray}
Each dipole contribution has the following form:
\begin{eqnarray}
{\cal D}_{ij,k} 
& = & 
- \frac{1}{2 p_i \cdot p_j}
{\cal A}_{5}^{(0)\;\ast}\left( p_1, ..., \tilde{p}_{(ij)},...,\tilde{p}_k,...\right)
\frac{{\bf T}_k \cdot {\bf T}_{ij}}{{\bf T}^2_{ij}} V_{ij,k} 
{\cal A}_{5}^{(0)}\left( p_1, ..., \tilde{p}_{(ij)},...,\tilde{p}_k,...\right).
 \nonumber
\end{eqnarray}
Here ${\bf T}_i$ denotes the colour charge operator for parton $i$
and $V_{ij,k}$ is a matrix in the spin space of the emitter parton
$(ij)$.
The momenta $ \tilde{p}_{(ij)}$ and $\tilde{p}_k$ are obtained from 
the momenta $p_i, p_j$ and $p_k$.
In general, the operators ${\bf T}_i$ lead to colour correlations, 
while the $V_{ij,k}$'s may lead to spin correlations.
The approximation $d\sigma^{\mathrm{A}}$ has to fulfill the requirement that
$d\sigma^{\mathrm{A}}$ is a proper approximation of $d\sigma^{\mathrm{R}}$ with
the same point-wise singular behaviour 
(in $D=4-2\epsilon$ dimensions) as
$d\sigma^{\mathrm{R}}$ itself.  Thus, $d\sigma^{\mathrm{A}}$ acts as a local counterterm for
$d\sigma^{\mathrm{R}}$, and one can safely perform the limit $\varepsilon
\rightarrow 0$. This defines the finite contribution
\begin{eqnarray}
\langle O \rangle^{\NLO}_{\{n+1\}} & = & 
 \int\limits_{n+1} \left( \left. O_{n+1} d\sigma^{\mathrm{R}} \right|_{\varepsilon=0} 
 - \left. O_n d\sigma^{\mathrm{A}} \right|_{\varepsilon=0} \right)\,.
\end{eqnarray}
The subtraction term can be integrated over the unresolved one-parton phase space.
Due to this integration, all spin correlations average out, 
but colour correlations still remain.
In a compact notation, the result of this integration is often written as
\begin{eqnarray}
 d\sigma^{\mathrm{C}} + \int\limits_1 d\sigma^{\mathrm{A}}  
 & = & {\bf I} \otimes d\sigma^B + {\bf K} \otimes d\sigma^B + {\bf P} \otimes d\sigma^B.
\end{eqnarray}
The notation $\otimes$ indicates that colour correlations still remain
and that an integration is involved.
The term ${\bf I} \otimes d\sigma^B$ lives on the phase space of the 
Born configuration and has the appropriate singularity structure to
cancel the IR divergences coming from the one-loop amplitude.
Therefore, $d\sigma^{\mathrm{V}} + {\bf I} \otimes d\sigma^B$
is IR finite.
The terms $({\bf K} + {\bf P} )\otimes d\sigma^B$ involve in addition 
an integration over the momentum fraction $x$ that rules the
collinear splitting of the incoming parton.
From the integration of the subtraction terms we obtain the finite
contribution
\begin{eqnarray}
  \langle O \rangle^{\NLO}_{\{n\}} & = & 
  \int\limits_n O_n \left( 
    d\sigma^{\mathrm{V}} 
    + {\bf I} \otimes d\sigma^B 
    + {\bf K} \otimes d\sigma^B 
    + {\bf P} \otimes d\sigma^B 
  \right)_{\varepsilon=0}\,.
\end{eqnarray}
The final structure of an NLO calculation in the subtraction formalism is then
\begin{eqnarray}
  \label{NLOcrosssection}
  \langle O \rangle^{\NLO} & = & 
  \langle O \rangle^{\NLO}_{\{n+1\}} 
  + \langle O \rangle^{\NLO}_{\{n\}}.
\end{eqnarray}
Since both contributions on the right-hand side of
\refeq{NLOcrosssection} are now finite, they can be evaluated with
numerical methods.  
The explicit forms of the dipole terms ${\cal D}_{ij,k}$,
together with the integrated counterparts, can be found in 
\citere{Catani:1996vz} for massless QCD and in
\citeres{Phaf:2001gc,Catani:2002hc} including massive quarks.

Analogously to our evaluation of the virtual corrections, we have also
performed two independent calculations of the real corrections.

One calculation of the real corrections 
results from a fully automated calculation
based on helicity amplitudes, as
described in \citere{Weinzierl:2005dd}.
Individual helicity amplitudes are computed with the help of
Berends--Giele recurrence relations \cite{Berends:1987me}.
The evaluation of colour factors and the generation of subtraction
terms is automated.
For the channel $\Pg \Pg \to \Pt \bar \Pt \Pg \Pg$ a dedicated
soft-insertion routine \cite{Weinzierl:1999yf} is used for the generation 
of the phase space.

The second calculation uses for the LO $2 \to 3$ processes
and the $\Pg \Pg \to \Pt \bar \Pt \Pg \Pg$ process optimized code obtained from
a Feynman diagrammatic approach.  As in the calculation described
before, standard techniques
like colour decomposition and the use of helicity amplitudes are
employed. For the $2 \to 4$ processes including light quarks, {\sl
  Madgraph} \cite{Stelzer:1994ta} has been used. The subtraction terms
according to \citere{Catani:2002hc} are obtained in a semi-automatized
manner based on a library written in {\sl C++}.

The two independent computer codes were compared point-wise at a few
phase-space points. In addition the entire numerical integration of the real 
corrections was done independently using the two codes. We found
complete agreement of the numerical results when the numerical uncertainty from 
the phase-space integration is taken into account. 

\section{Numerical results}
\label{sect:results}

\subsection{Setup}

In the following we consistently use the CTEQ6~\cite{Pumplin:2002vw}
set of parton distribution functions (PDFs). In detail, we take
CTEQ6L1 PDFs with a one-loop running $\alpha_{\mathrm{s}}$ in
LO and CTEQ6M PDFs with a two-loop running $\alpha_{\mathrm{s}}$
in NLO.
The number of active flavours is $N_{\mathrm{F}}=5$, and the
respective QCD parameters are $\Lambda_5^{\mathrm{LO}}=165\MeV$
and $\Lambda_5^{\overline{\mathrm{MS}}}=226\MeV$.
As mentioned earlier the top-quark loop in the gluon self-energy is
subtracted at zero momentum. In this scheme the running of
$\alpha_{\mathrm{s}}$ is generated solely by the contributions of the
light quark and gluon loops. The top-quark mass is
renormalized in the on-shell scheme, as numerical value we take $\Mt=174\GeV$.
If not stated otherwise, we identify the renormalization and factorization
scales, $\mu_{\mathrm{ren}}$ and $\mu_{\mathrm{fact}}$, with $\Mt$.

For the definition of the tagged hard jet we apply the jet algorithm 
of \citere{Ellis:1993tq} with $R=1$  and
require a transverse momentum of
$\ptjet>\ptjetcut$ with $\ptjetcut=20\GeV$ and $\ptjetcut=50\GeV$ for
the hardest jet at the Tevatron and the LHC, respectively.
The outgoing (anti)top-quarks are neither affected
by the jet algorithm nor by the phase-space cut. We assume them as
always tagged.
Note that the LO prediction and the virtual corrections are not influenced
by the recombination procedure of the jet
algorithm, but the real corrections are.

Up to the transverse-momentum cut $\ptjetcut=50\GeV$ for the
LHC, the setup used in this article coincides with the one used in 
\citere{Dittmaier:2007wz}.
There, $\ptjetcut=20\GeV$ was used both for the Tevatron and the LHC.

\subsection{Results for the Tevatron}

As discussed in \citere{Dittmaier:2007wz}, the integrated LO 
cross section for $\Pt \bar \Pt + \onejet $ production at 
the Tevatron is dominated by the $q\bar q$ channel 
with about 85\%, followed by the $\Pg\Pg$ channel with about 7\%.
This is rather similar to inclusive top-quark pair cross section
where again at LO about 90\% is obtained from $q\bar q$ channel and about 10\%
from the $\Pg\Pg$ channel. In difference to the inclusive case the $q\Pg$
($\bar q\Pg$) are not suppressed in the coupling. This accounts for the
slightly larger contribution from these channels.  One should keep in
mind that the precise contribution of individual channels depends on
the factorisation scale as well as on the chosen parton distributions.
As a consequence the aforementioned numbers give
just a qualitative picture.

\begin{table}
  \begin{center}
    \leavevmode
    \begin{tabular}{c|l|l|l|l}
      &\multicolumn{2}{|c|}{$\sigma_{\Pt\bar\Pt\mathrm{jet}}[\pba]$}
      &\multicolumn{2}{|c}{$A^{\Pt}_{\mathrm{FB}}[\%]$}\\
      $\ptjetcut$ [GeV]&\multicolumn{1}{|c|}{LO}&
      \multicolumn{1}{|c|}{NLO}&
      \multicolumn{1}{|c|}{LO}
      &\multicolumn{1}{|c}{NLO}\\ \hline 
      20& 1.583(2)$^{+0.96}_{-0.55}$ & 1.791(1)$^{+0.16}_{-0.31}$ &
      $-7.69(4)^{+0.10}_{-0.085}$ & $-1.77(5)^{+0.58}_{-0.30}$ \\

      30& 0.984(1)$^{+0.60}_{-0.34}$ & 1.1194(8)$^{+0.11}_{-0.20}$ &
      $-8.29(5)^{+0.12}_{-0.085}$ & $-2.27(4)^{+0.31}_{-0.51}$ \\

      40& 0.6632(8)$^{+0.41}_{-0.23}$ & 0.7504(5)$^{+0.072}_{-0.14}$ &
      $-8.72(5)^{+0.13}_{-0.10}$ & $-2.73(4)^{+0.35}_{-0.49}$ \\

      50& 0.4670(6)$^{+0.29}_{-0.17}$ & 0.5244(4)$^{+0.049}_{-0.096}$
      & $-8.96(5)^{+0.14}_{-0.11}$ & $-3.05(4)^{+0.49}_{-0.39}$\\
      \hline 
\end{tabular}
\caption{Cross section $\sigma_{\Pt\bar\Pt\mathrm{jet}}$
  and forward--backward charge asymmetry $A^{\Pt}_{\mathrm{FB}}$ at the
  Tevatron for different values of $\ptjetcut$ 
  for $\mu=\mu_{\mathrm{fact}}=\mu_{\mathrm{ren}} = \Mt$.
  The upper and lower indices are the shifts towards $\mu = \Mt/2$
  and $\mu = 2\Mt$.}
\label{tab:tevcs}
\end{center}
\end{table}
In \refta{tab:tevcs} we provide the LO and NLO predictions for the
integrated cross sections for different values of the cut on the transverse
momentum of the hard jet (left part).
The values presented are for the central scale 
$\mu=\mu_{\mathrm{fact}}=\mu_{\mathrm{ren}}=\Mt$.
In parentheses we quote the uncertainty due to the numerical
integration. The scale dependence is indicated by the upper and lower
indices. The upper (lower) index represents the change when the scale
is shifted towards $\mu = \Mt/2$ ($\mu = 2\Mt$). 
Rescaling the common scale $\mu=\mu_{\mathrm{fact}}=\mu_{\mathrm{ren}}$ 
from the default value
$\Mt$ up (down) by a factor 2 changes the cross section in LO and NLO
by about $60\%$ ($35\%$) and $9\%$ ($18\%$), respecively, i.e.\
the scale uncertainty is reduced considerably through the inclusion of 
the NLO corrections. The above findings are rather insensitive to the
chosen cut value. We find only variations at the per-cent level. In
particular, there is no big difference for the lowest cut value
compared to the other values, suggesting that the perturbative
expansion is under control and not
spoiled by the appearance of large logarithms. 
Compared with the total cross section we find
that for the small $p_\rT$ cut of 20 $\GeV$ the $\Pt\bar\Pt +\onejet$
events represent almost 30 \% of the total cross section. This
fraction shows an evident dependence on the value chosen for the $p_\rT$
cut. The fraction is reduced to about 8\% when 50 $\GeV$ is chosen
for the cut. 
The NLO corrections change the ratio for a given value of $p_\rT$ 
cut only at the level of a few per cent.

In the right part of \refta{tab:tevcs} we show results 
for the forward--backward charge asymmetry.
In LO the top-quark charge asymmetry is defined by
\begin{equation}
  \label{eq:asym}
  A^{\Pt}_{\mathrm{FB,LO}} =
  \frac{\sigma^-_{\mathrm{LO}}}{\sigma^+_{\mathrm{LO}}},
\end{equation}
with the definition
\begin{eqnarray}
  \sigma^\pm_{\mathrm{LO}} =
  \sigma_{\mathrm{LO}}(y_{\Pt}{>}0)\pm\sigma_{\mathrm{LO}}(y_{\Pt}{<}0),
\end{eqnarray}
where $y_{\Pt}$ denotes the rapidity of the top-quark.
Cross-section contributions
$\sigma(y_{\Pt}$ \raisebox{-.2em}{$\stackrel{>}{\mbox{\scriptsize$<$}}$} $0)$
correspond to top-quarks in the forward or backward hemispheres, respectively,
where incoming protons fly into the forward direction by definition.
Denoting the corresponding NLO contributions to the cross sections by
$\delta\sigma^\pm_{\mathrm{NLO}}$,
we define the asymmetry at NLO by
\begin{equation}
\label{eq:NLOasym}
 A^{\Pt}_{\mathrm{FB,NLO}} =
\frac{\sigma^-_{\mathrm{LO}}}{\sigma^+_{\mathrm{LO}}}
\left( 1+
 \frac{\delta\sigma^-_{\mathrm{NLO}}}{\sigma^-_{\mathrm{LO}}}
-\frac{\delta\sigma^+_{\mathrm{NLO}}}{\sigma^+_{\mathrm{LO}}} \right),
\end{equation}
i.e.\ via a consistent expansion in $\alpha_{\mathrm{s}}$.
Note, however, that the LO cross sections in \refeq{eq:NLOasym}
are evaluated in the NLO setup (PDFs, $\alpha_{\mathrm{s}}$).
In \citere{Dittmaier:2007wz} it was already pointed out that
the LO asymmetry for a $p_{\rT}$-cut of 20 $\GeV$, 
which is about $-7.7\%$ with a small scale
uncertainty, is reduced to about $-1.8\%$ with the rather large
scale uncertainty that is---assessed conservatively---almost as large
as its absolute size. The reason for this growing scale uncertainty
when going from LO to NLO simply results from the fact that the LO
prediction for $A^{\Pt}_{\mathrm{FB}}$ is independent of the 
renormalization scale, since the strong coupling drops out in the ratio.
Thus, the scale dependence does not reflect the total theoretical
uncertainty in LO at all for this quantity.
Table~\ref{tab:tevcs} shows that this feature qualitatively holds true
also for larger values of $\ptjetcut$ used for the jet definition,
but the LO and NLO asymmetries are shifted towards larger absolute
values for a larger cut.

\begin{figure}
\epsfig{file=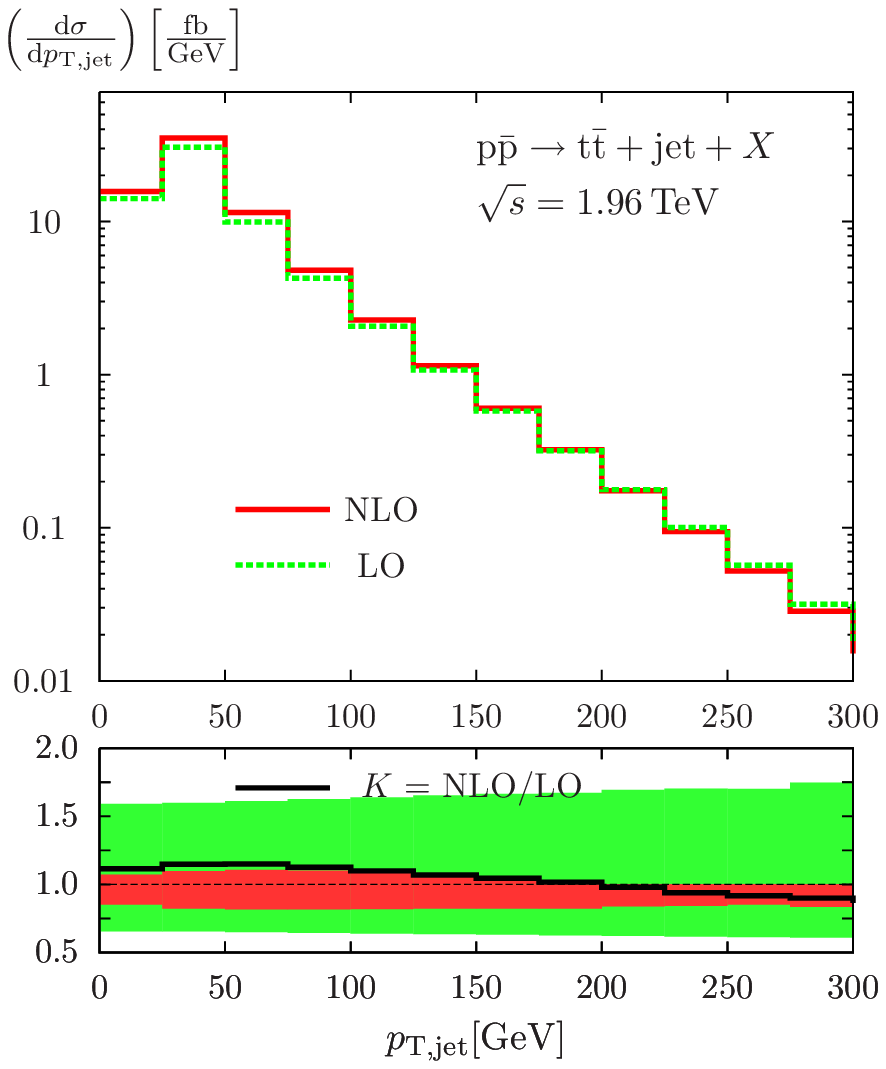,width=.48\textwidth}
\epsfig{file=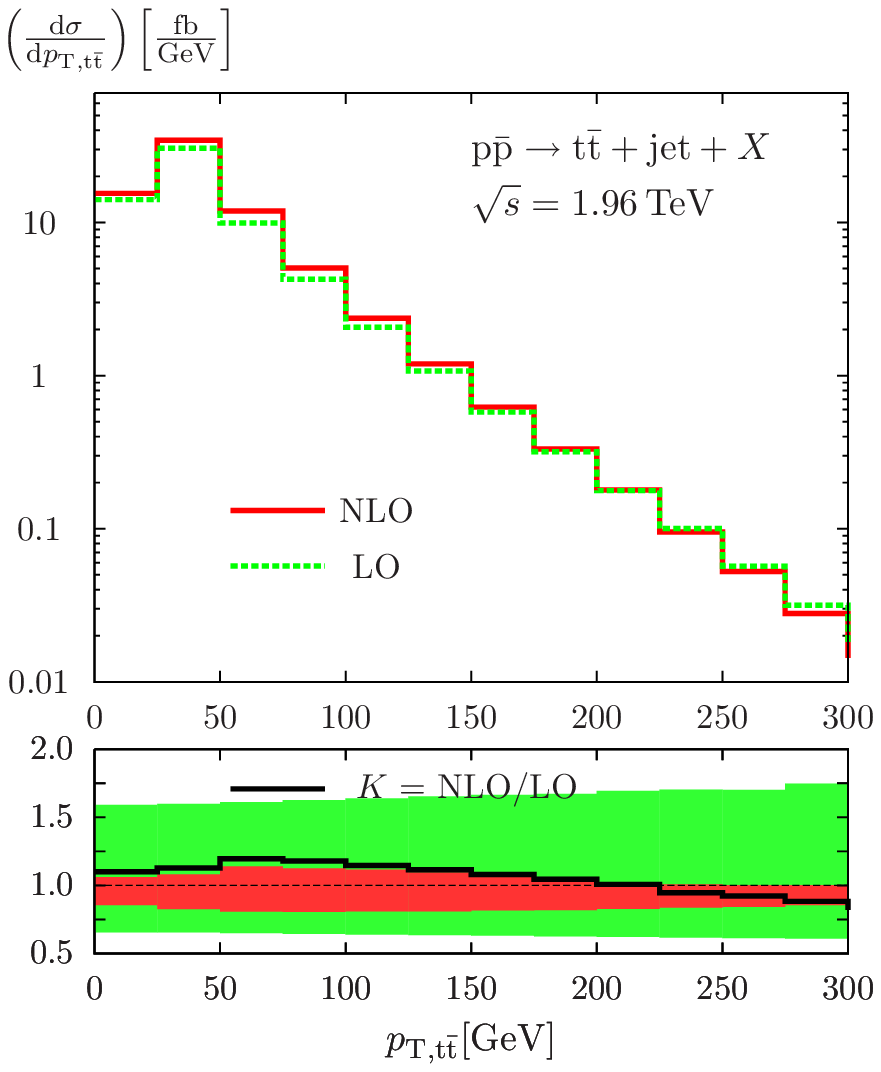,width=.48\textwidth}
\\[1em]
\epsfig{file=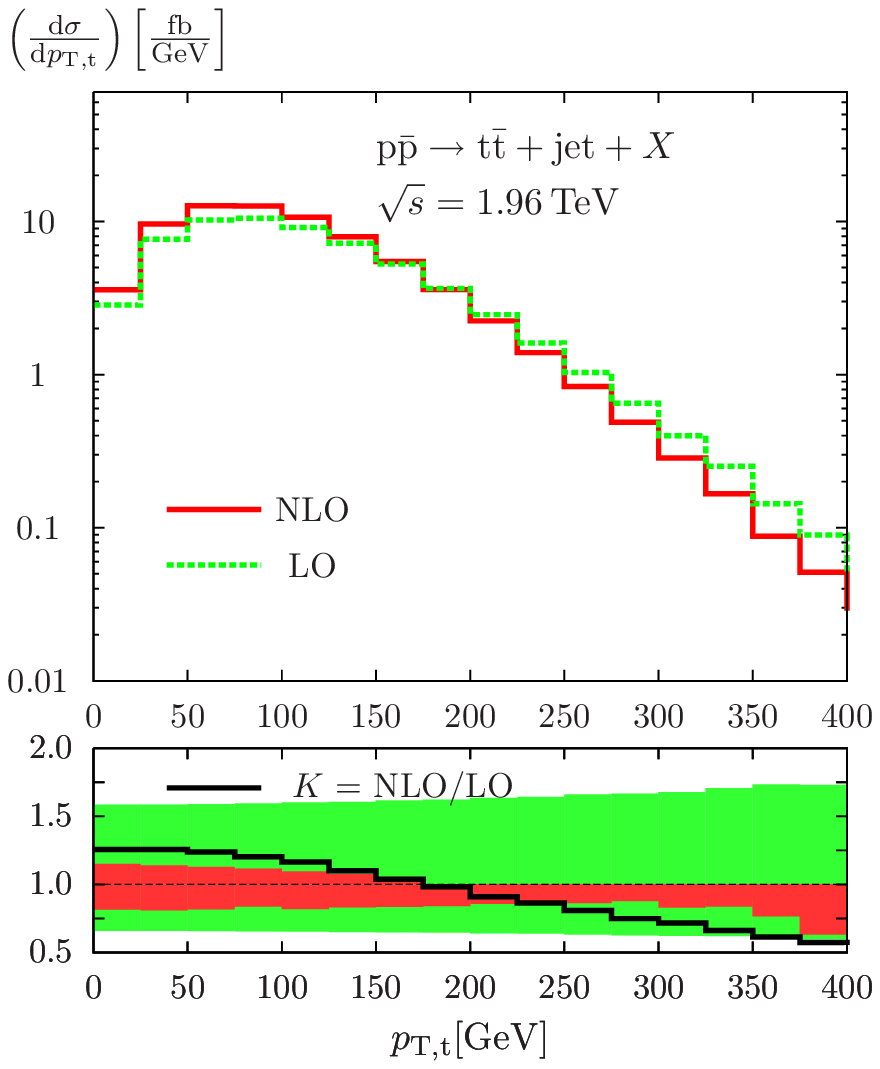,width=.48\textwidth}
\caption{Transverse-momentum distributions of the hard jet
($\ptjet$), of the total $\Pt\bar\Pt$ system ($p_{\rT,\Pt\bar\Pt}$),
and of the top-quark ($p_{\rT,\Pt}$) at the Tevatron.
The lower panels show the ratios $K=\mathrm{NLO/LO}$ as well as the
LO and NLO scale uncertainties corresponding to a rescaling of
$\mu=\mu_{\mathrm{fact}}=\mu_{\mathrm{ren}} = \Mt$ by a factor 2.}
\label{fig:pt_tev}
\end{figure}

Figure~\ref{fig:pt_tev} shows the distributions in the transverse
momenta of the hard jet,
$\ptjet$, of the total $\Pt\bar\Pt$ system, $p_{\rT,\Pt\bar\Pt}$,
and of the top-quark, $p_{\rT,\Pt}$.
In LO $\ptjet$ and $p_{\rT,\Pt\bar\Pt}$ coincide because of momentum
conservation in the transverse plane, but the radiation of two jets
in the real corrections renders them different. The numerical results
in \reffi{fig:pt_tev}, however, reveal that the differences are very small.
The shown $p_{\rT}$ distributions drop with growing $p_{\rT}$, where
the spectrum for the top-quark is much harder than the ones for the
jet and the $\Pt\bar\Pt$ system. 
In the $\ptjet$ and $p_{\rT,\Pt\bar\Pt}$ spectra 93\% of the events 
are concentrated below a $p_\rT$ of about $100\GeV$ at NLO, while 92\%
of the events have a $p_{\rT,\Pt}$ with less than  $200\GeV$.
Employing a fixed scale $\mu=\Mt$,
the NLO corrections do not simply rescale the LO shape, but induce
distortions at the level of some 10\%, which redistribute events
from larger to smaller transverse momenta. 
We believe that two effects contribute to these distorsions.
First of all, the use of a fixed renormalization scale $\mu_{\mathrm{ren}}=\Mt$ 
is not an appropriate choice for high $p_{\rT}$ events.
Due to the large value of $\alpha_{\mathrm{s}}$ the LO calculation overestimates
the cross section for high $p_{\rT}$ and the NLO calculation has to compensate this
scale choice. 
We expect that the distortions due to this effect are  reduced if an appropriate $p_{\rT}$-dependent
scale choice is used, such as $\mu=\sqrt{p_{\rT}^2+\Mt^2}$. 
As a second effect in particular the $p_{\rT}$ distribution of the top-quarks
can become softer due to the emission of additional particles, which is accounted for the first time
by an NLO calculation.

As can also be seen from the lower panel of each plot, we find again
an important reduction of the scale dependence when the NLO
corrections are taken into account. At least for the $\ptjet$- and 
$p_{\rT,\Pt\bar\Pt}$-distribution the corrections are of moderate
size. For the $p_{\rT,\Pt}$-distribution we find large corrections for 
large values of $p_{\rT,\Pt}$. The corrections are almost 50\% for a
$p_\rT$ arround 400 $\GeV$. As mentioned this could probably be cured by 
employing a $p_{\rT}$-dependent scale.

\begin{figure}
  \epsfig{file=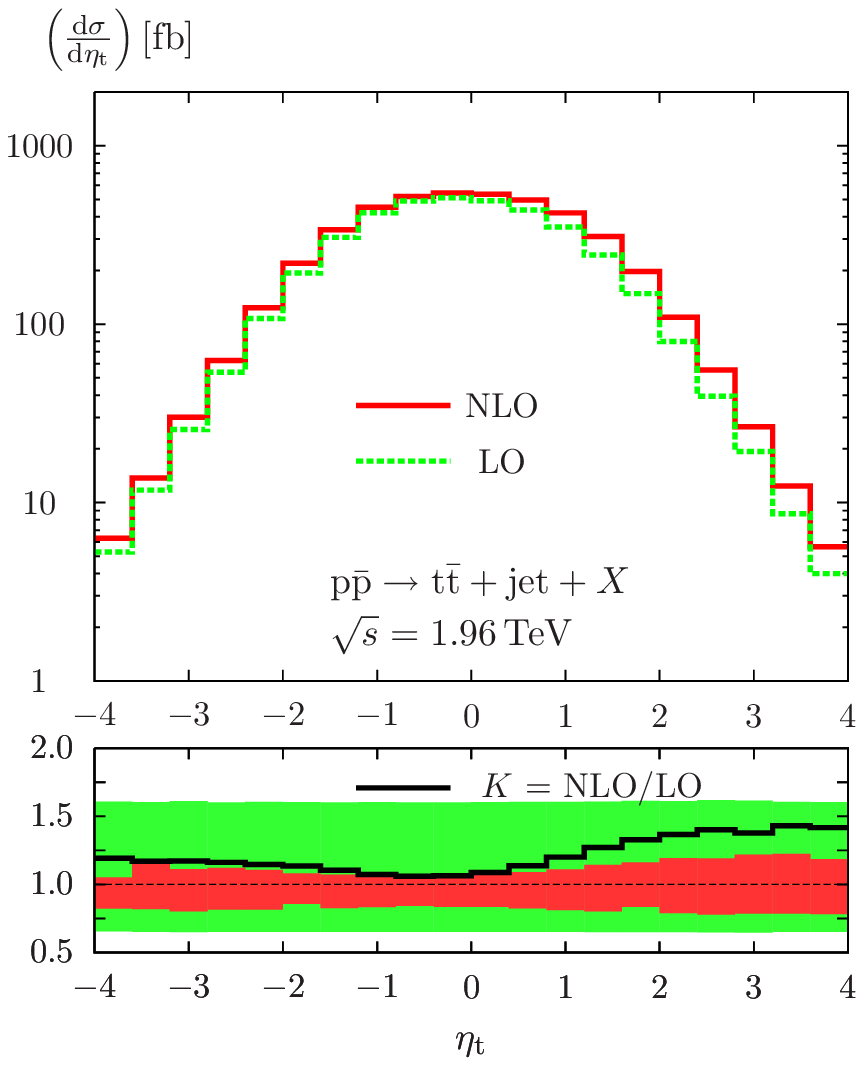,width=.47\textwidth}
  \epsfig{file=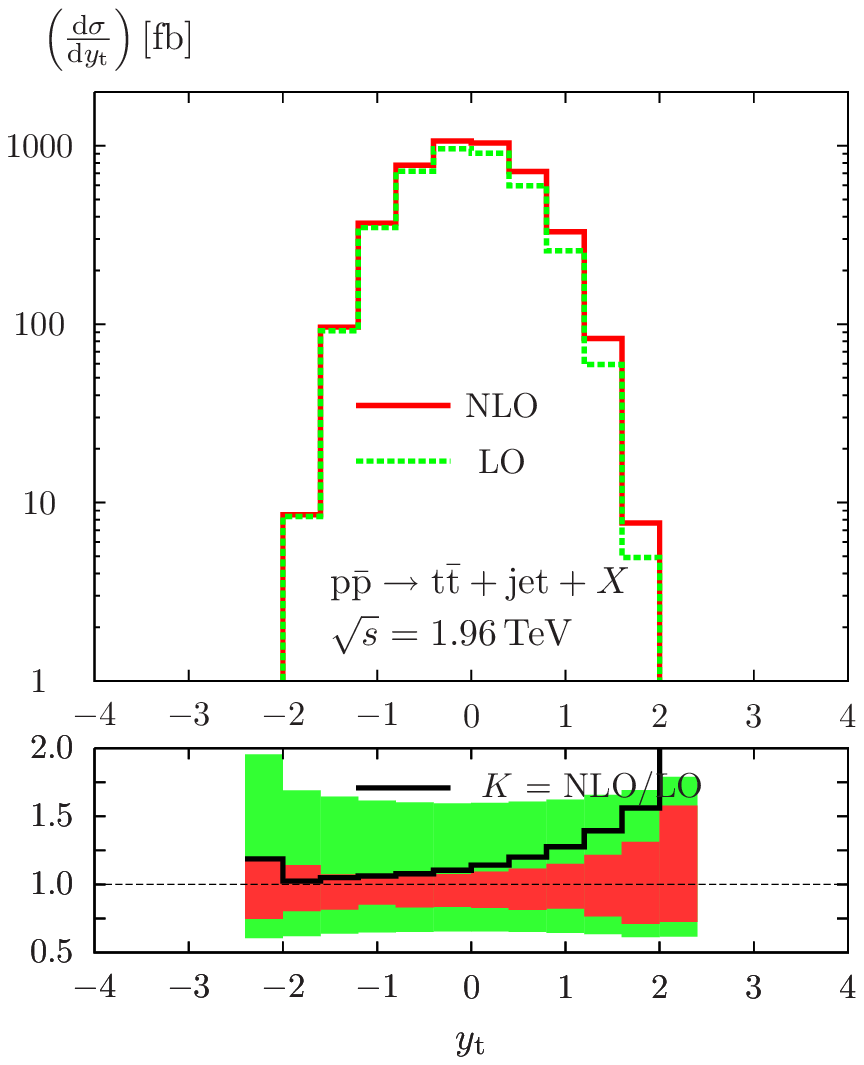,width=.47\textwidth}
  \\[1em]
  \epsfig{file=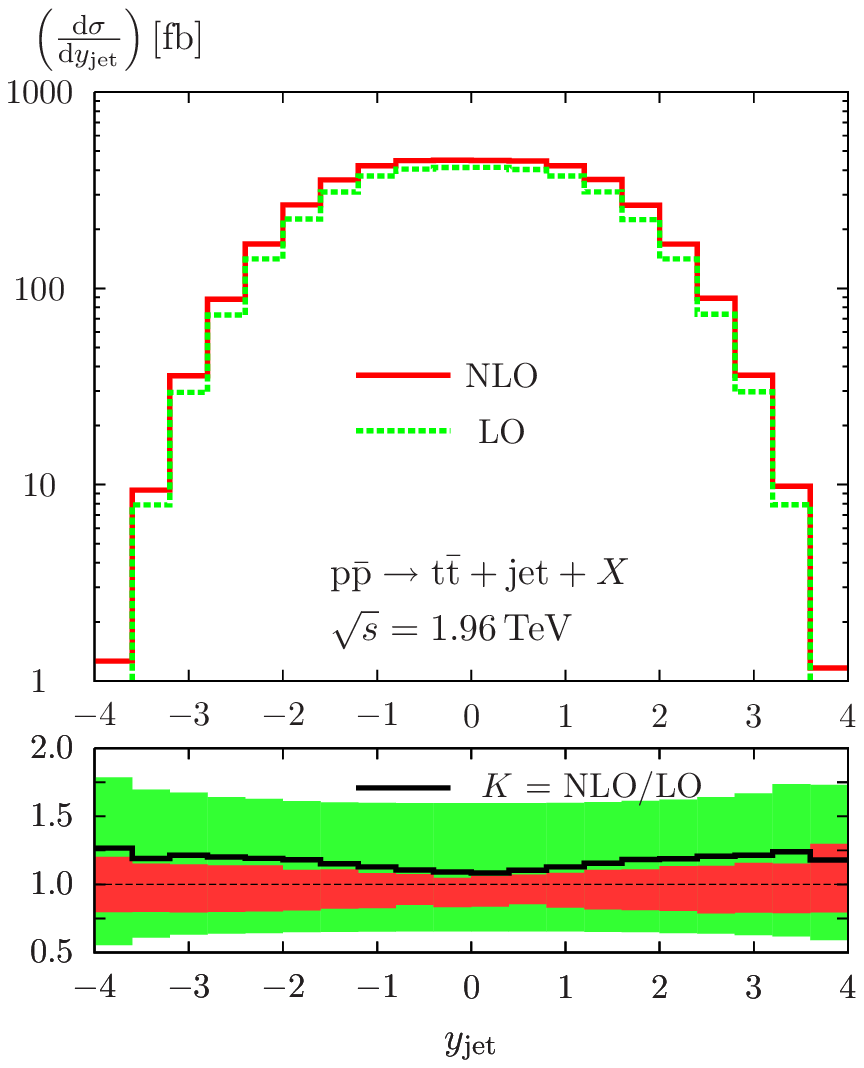,width=.47\textwidth}
  \caption{Distributions in the pseudo-rapidity ($\eta_\Pt$) and 
    rapidity ($y_\Pt$) of the top-quark, and in the rapidity
    ($y_{\mathrm{jet}}$) of the hard jet at the Tevatron.
    The lower panels show the ratios $K=\mathrm{NLO/LO}$ as well as the
    LO and NLO scale uncertainties corresponding to a rescaling of
    $\mu=\mu_{\mathrm{fact}}=\mu_{\mathrm{ren}} = \Mt$ by a factor 2.}
  \label{fig:y_tev}
\end{figure}
Figure~\ref{fig:y_tev} depicts the distributions in the pseudo-rapidity 
and rapidity of the top-quark, $\eta_\Pt$ and $y_\Pt$, and in the rapidity,
$y_{\mathrm{jet}}$, of the hard jet. For massless momenta the
pseudo-rapidity, which is defined through
\begin{equation}
  \eta = -\ln\left(\tan\left({\vartheta\over 2}\right)\right)
\end{equation}
($\vartheta$ is the scattering angle with respect to the beam axis),
is equivalent to the rapidity defined through
\begin{equation}
  y = {1\over 2} \ln\left({ E+p_z\over E-p_z }\right)
\end{equation}
($E$ denotes the energy, $p_z$ the three-momentum component along the 
beam axis). 
For the massive top-quark we observe a rather important difference between the
$\eta_\Pt$ and $y_\Pt$ distribution.
Recently there has been significant interest in the rapidity distribution of the
jet, as MC@NLO \cite{Frixione:2003ei} and Alpgen with MLM matching \cite{Mangano:2006rw} disagree
on this distribution.
Our result includes for the first time the full ${\cal O}(\alpha_s^4)$
matrix elements.
Note that the set-up of ref.~\cite{Mangano:2006rw} differs from the one used
here, therefore the distributions should not be compared directly.

At NLO, 90\% of the events are concentrated within
$|\eta_\Pt| < 2.0$. Demanding $|y_\Pt| < 1.2$ selects 96\% of the
events. For $y_{\mathrm{jet}}$ we find that  94\% of the events are
contained in $|y_{\mathrm{jet}}| < 2.4$.

The reduction of the forward--backward asymmetry $A^{\Pt}_{\mathrm{FB}}$
discussed above induced by the NLO corrections is clearly visible
in the $\eta_\Pt$ and $y_\Pt$ distributions. The corrections are
larger in the forward direction. The asymmetry in the LO distributions
is thus reduced by the NLO corrections. It is hardly conceivable
that this higher-order effect can be absorbed into LO predictions
by phase-space-dependent scale choices.
It should be realized that the forward--backward-symmetric rapidity 
distribution of the hard jet gets distorted by the corrections as
well. The corrections increase for large values of $|y_{\mathrm{jet}}|$.

At least in the regions of the distributions in which the rate is not
too much suppressed, the NLO corrections reduce the scale uncertainty of the
LO distributions in a similar way as observed for the integrated cross section.

\subsection{Results for the LHC}

Table~\ref{tab:lhccs} shows the integrated cross section for various
values of the cut $\ptjetcut$ on the transverse momentum of the hard
tagging jet at the LHC.
\begin{table}
  \begin{center}
    \leavevmode
    \begin{tabular}{c|l|l} \multicolumn{1}{c}{}
      &\multicolumn{2}{c}{$\sigma_{\Pt\bar\Pt\mathrm{jet}}[\pba]$} \\
      $\ptjetcut$ [GeV]&\multicolumn{1}{|c|}{LO}
      &\multicolumn{1}{|c}{NLO}\\ \hline 
      20 & 710.8(8)$^{+358}_{-221}$ & 692(3)3$^{-40}_{-62}$ \\
      50 & 326.6(4)$^{+168}_{-103}$ & 376.2(6)$^{+17}_{-48}$ \\
      100& 146.7(2)$^{+77}_{-47}$   & 175.0(2)$^{+10}_{-24}$ \\
      200& 46.67(6)$^{+26}_{-15}$   & 52.81(8)$^{+0.8}_{-6.7}$ \\
      \hline 
\end{tabular}
\caption{Cross section $\sigma_{\Pt\bar\Pt\mathrm{jet}}$
at the LHC for different values of $\ptjetcut$ 
      for $\mu=\mu_{\mathrm{fact}}=\mu_{\mathrm{ren}} = \Mt$.
      The upper and lower indices are the shifts towards $\mu = \Mt/2$
      and $\mu = 2\Mt$.}
    \label{tab:lhccs}
  \end{center}
\end{table}
In contrast to the Tevatron, the $\Pg\Pg$ channel comprises about
70\% of the LO $\Pp\Pp$ cross section, followed by $q\Pg$ with about 
22\%~\cite{Dittmaier:2007wz}. We note that the importance of the $q\Pg$
channel is very different from the inclusive top-quark
production. For inclusive top-quark pair production this channel is 
suppressed---despite the large parton luminosity in this
channel---because it appears only at NLO. For $\Pt \bar \Pt + \onejet$ 
production the $q\Pg$ channel appears already in LO and thus gives a
significant contribution due to the large parton luminosity.
Comparing the LO and NLO predictions we find again that
the large scale dependence of about 100\% in the LO cross section
is considerably reduced after including the NLO corrections.
The ratio of the NLO $\Pt\bar\Pt +\onejet$ cross section to the total
NLO $\Pt\bar\Pt$ cross section is about 47\%, 22\%, and 7\% for a
$p_\rT$ cut of $50\GeV$, $100\GeV$, and $200\GeV$, respectively. 

\begin{figure}
  \epsfig{file=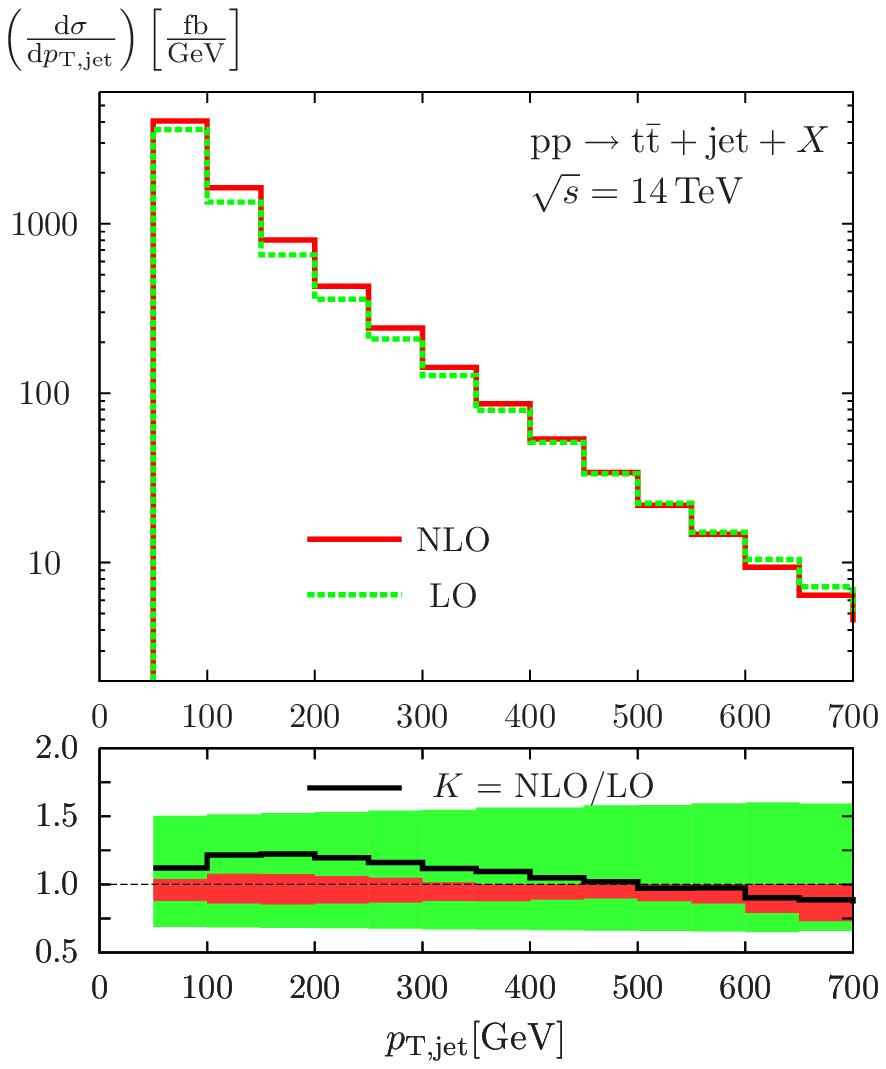,width=.48\textwidth}
  \epsfig{file=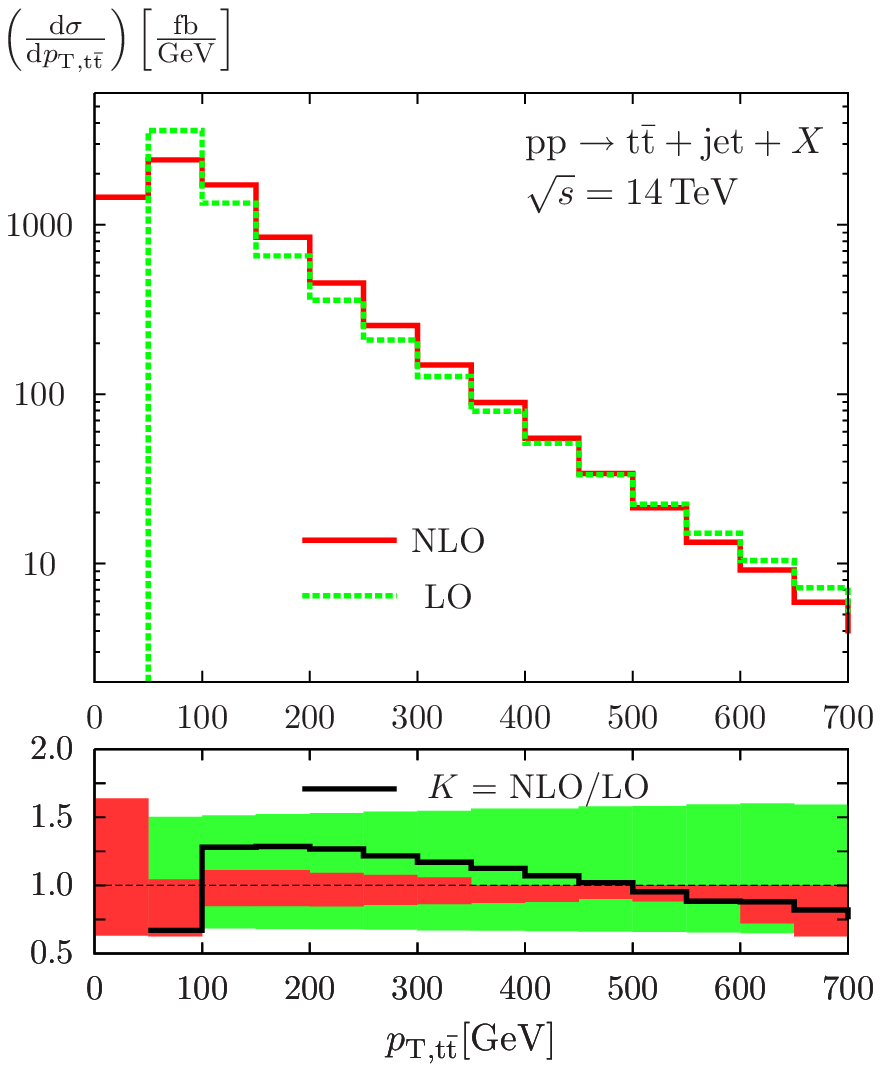,width=.48\textwidth}
  \\[1em]
  \epsfig{file=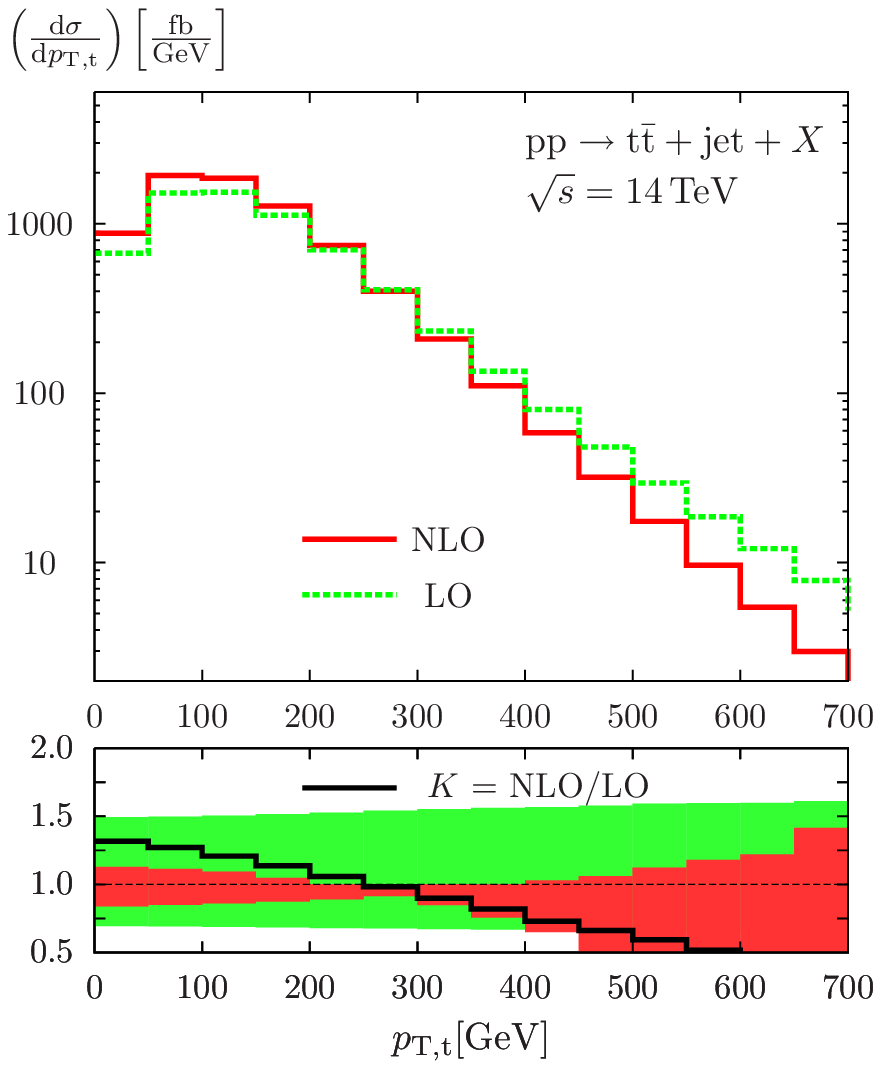,width=.48\textwidth}
  \caption{Transverse-momentum distributions of the hard jet
    ($\ptjet$), of the total $\Pt\bar\Pt$ system ($p_{\rT,\Pt\bar\Pt}$),
    and of the top-quark ($p_{\rT,\Pt}$) at the LHC.
    The lower panels show the ratios $K=\mathrm{NLO/LO}$ as well as the
    LO and NLO scale uncertainties corresponding to a rescaling of
    $\mu=\mu_{\mathrm{fact}}=\mu_{\mathrm{ren}} = \Mt$ by a factor 2.}
  \label{fig:pt_lhc}
\end{figure}

In \reffi{fig:pt_lhc} we show the distributions in the transverse
momenta of the hard jet,
$\ptjet$, of the total $\Pt\bar\Pt$ system, $p_{\rT,\Pt\bar\Pt}$,
and of the top-quark, $p_{\rT,\Pt}$.
The distributions become harder in $p_{\rT}$ when going from the
Tevatron to the LHC, as expected from the higher scattering energy.
At NLO, 92\% of the events have transverse momenta
$\ptjet<250\GeV$,  and 94\% have
$p_{\rT,\Pt}<300\GeV$ in the respective distributions.
In contrast to the Tevatron, the $\ptjet$ and $p_{\rT,\Pt\bar\Pt}$ 
distributions, which are identical in LO become different in NLO.
For the $\ptjet$-distribution the lowest bin ($0<\ptjet<50\GeV$) 
is always empty
due to the cut applied. For the $p_{\rT,\Pt\bar\Pt}$ this holds also
true in LO because the transverse momenta between the $\Pt\bar\Pt$
system and the additional hard jet are balanced. In NLO the lowest bin
in the $p_{\rT,\Pt\bar\Pt}$ distribution is populated due to
an additional jet.
For large $p_{\rT}$ the difference between $\ptjet$ and $p_{\rT,\Pt\bar\Pt}$ 
distributions is at the level of about 10\%. This is again due to the
presence of the additonal jet.
As observed already for the Tevatron, the shapes of the $p_{\rT}$
distributions receive distortions by the corrections.

\begin{figure}
  \epsfig{file=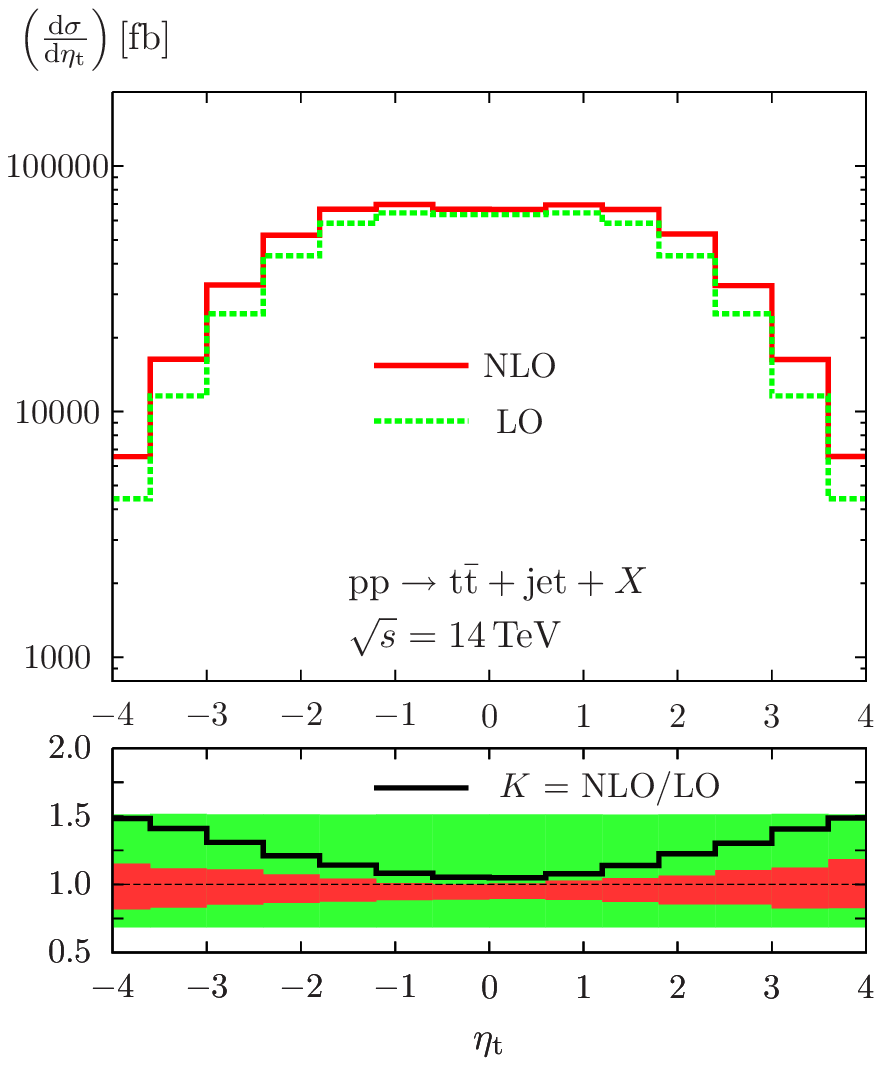,width=.48\textwidth}
  \epsfig{file=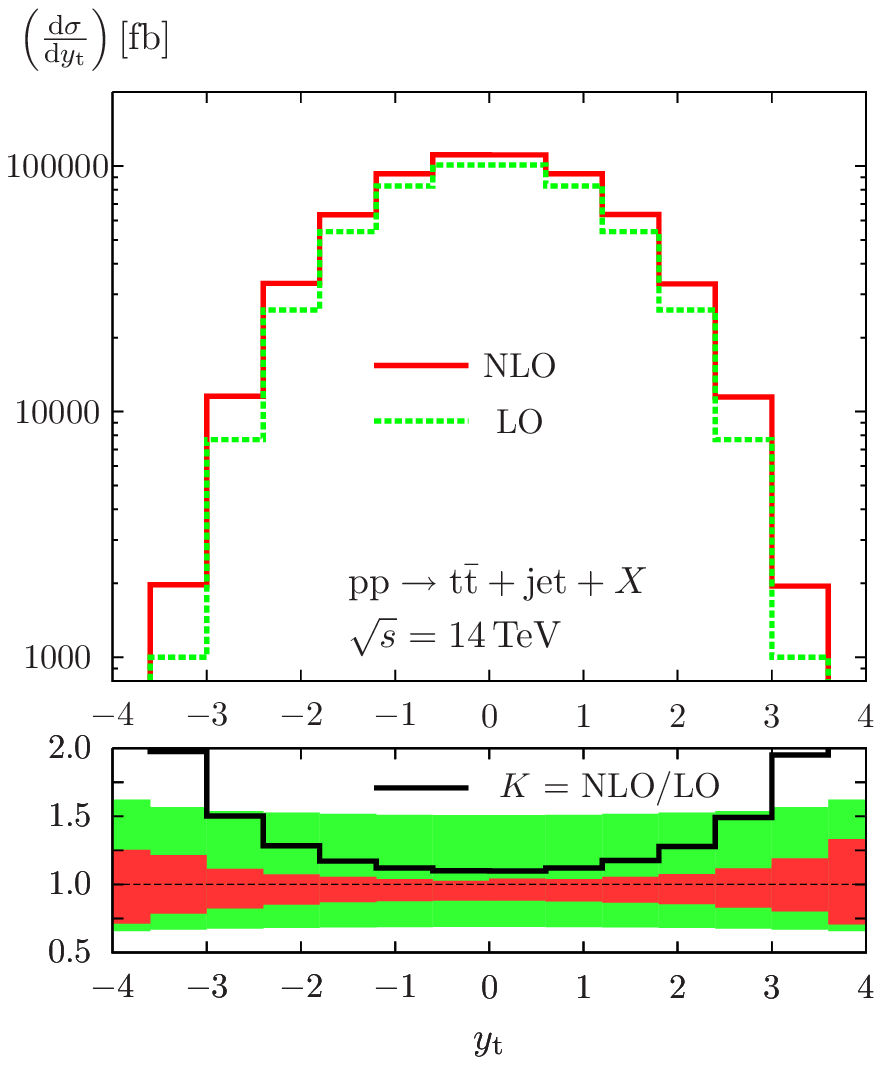,width=.48\textwidth}
  \\[1em]
  \epsfig{file=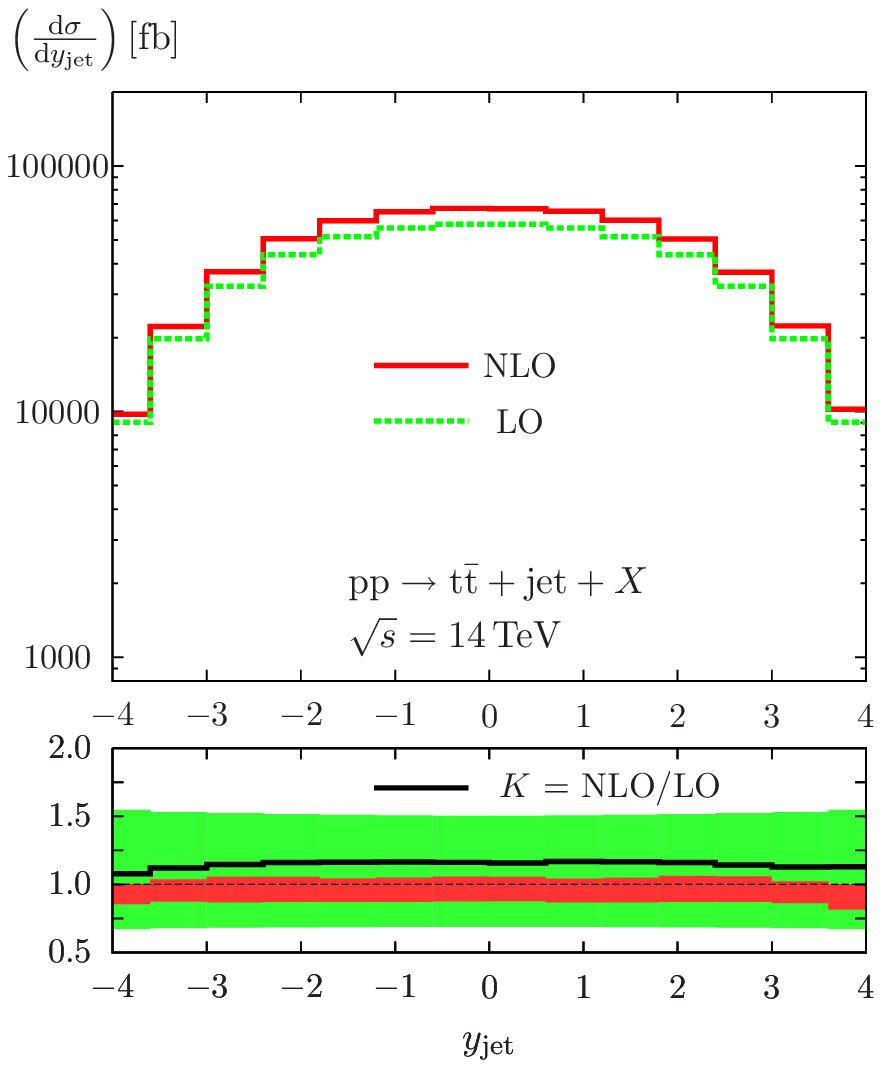,width=.48\textwidth}
  \caption{Distributions in the pseudo-rapidity ($\eta_\Pt$) and 
    rapidity ($y_\Pt$) of the top-quark, and in the rapidity
    ($y_{\mathrm{jet}}$) of the hard jet at the LHC.
    The lower panels show the ratios $K=\mathrm{NLO/LO}$ as well as the
    LO and NLO scale uncertainties corresponding to a rescaling of
    $\mu=\mu_{\mathrm{fact}}=\mu_{\mathrm{ren}} = \Mt$ by a factor 2.}
  \label{fig:y_lhc}
\end{figure}

Figure~\ref{fig:y_lhc} illustrates the distributions in the pseudo-rapidity 
and rapidity of the top-quark, $\eta_\Pt$ and $y_\Pt$, and in the rapidity,
$y_{\mathrm{jet}}$, of the hard jet.
In the respective distributions 92\% of the NLO events concentrate within 
the regions $|\eta_\Pt| < 3.0$, 96\% have $|y_\Pt|< 2.4$, and 96\%
have $|y_{\mathrm{jet}}| < 3.6$ in the respective distributions, i.e.\
these distributions get broadened roughly by one unit in the
transistion from Tevatron to LHC.
As for the Tevatron, we find distortions of the shapes induced by the
corrections that are hard to mimic by phase-space-dependent scale
choices. All the shown $y$ and $\eta$ distributions at the LHC are
forward--backward-symmetric, but actually the distributions of top
and antitop-quarks are intrinsically different. Numerically we do not
observe a significant difference, so that we show only the
distributions for the top-quark.

Again, in all shown distributions a reduction of the scale uncertainty by
the NLO corrections is
visible that is comparable to the one in the integrated cross section.

\section{Conclusions}

The production of $\Pt\bar\Pt{+}$jet final states represents important
processes both at the Tevatron and the LHC. The signal is interesting
in its own right, because large fractions of the $\Pt\bar\Pt$ samples
show additional jet activity and deviations from the SM could signal
new physics such as top-quark compositeness. Moreover,
$\Pt\bar\Pt{+}$jet production delivers a large background to many
searches at the LHC, such as for the Higgs boson via 
weak-vector-boson fusion.

We have presented NLO QCD predictions for 
$\Pt\bar\Pt{+}$jet production at the Tevatron and the LHC.
The NLO corrections reduce the scale uncertainty of the total cross
section and of the differential distributions compared to a LO calculation,
which can only provide qualitative predictions.
Further theoretical improvements could only be achieved by dedicated
QCD resummations, since a full treatment at NNLO is certainly out of reach.
Already the presented NLO calculation is quite complicated.
For this reason we have also documented a set of numerical results
for the one-loop correction and for the real-emission parts at single 
phase-space points, in order to facilitate comparisons to our
calculation by other groups.

The charge asymmetry of the top-quark, which is measured at the
Tevatron, is significantly decreased at NLO and is almost washed out
by the residual scale dependence. We have studied 
the dependence of the NLO asymmetry on the cut on the transverse
momentum of the hard tagging jet. Further refinements
in the description of the charge asymmetry are required
to stabilize the predictions with respect to higher-order corrections.
Moreover, the top-quark decay should be taken into account properly.

Finally, the presented NLO QCD calculation for the $\Pt\bar\Pt{+}$jet 
process represents a building block for a full NNLO QCD prediction
for $\Pt\bar\Pt$ production, a demanding and important calculation
that is currently in progress by various groups.

\section*{Acknowledgements}
This work is supported in part by the European Community's Marie-Curie
Research Training Network under contract MRTN-CT-2006-035505 ``Tools
and Precision Calculations for Physics Discoveries at Colliders''
and by Deutsche Forschungsgemeinschaft (DFG)  through SFB/TR 9.
P.U. is financed through a Heisenberg fellowship of DFG.

\newpage

\appendix
\section*{Appendix}
\section{Benchmark numbers for the virtual corrections}

In order to facilitate a comparison to our calculation, 
we provide explicit numbers on the
squared LO amplitude and the corresponding virtual corrections for a
single non-exceptional phase-space point. The set of momenta for
$ab\to \Pt\bar\Pt c$ with the explicit partonic reactions 
$\Pg\Pg\to \Pt\bar\Pt\Pg$,
$q\bar q \to \Pt\bar\Pt\Pg$, 
$q\Pg \to \Pt\bar\Pt q$, and 
$\Pg\bar q \to \Pt\bar\Pt\bar q$ 
is chosen as
\begin{eqnarray}
p_a &=& \scriptstyle (500,0,0,500),
\nn\\
p_b &=& \scriptstyle (500,0,0,-500),
\nn\\
p_{\Pt} &=& \scriptstyle
(458.5331753852783,
 207.0255169909440,
 0,
 370.2932732896167),
\nn\\
p_{\bar\Pt} &=& \scriptstyle
(206.6000026080000,
 -10.65693677252589,
  42.52372780926147,
 -102.3998210421085),
\nn\\
p_c &=& \scriptstyle
( 334.8668220067217,
 -196.3685802184181,
  -42.52372780926147,
 -267.8934522475083),
\label{eq:PSpoint}
\end{eqnarray}
with the obvious notation
$p = (p^0,p^1,p^2,p^3)$
and all the components given in GeV. 
The top-quark mass is set to $\Mt=174\GeV$.
We give numbers on the spin- and colour-averaged squared LO amplitude 
$|{\cal A}_5^{(0)}|^2$ as well as  for the contribution
$2\mathrm{Re}({\cal A}_5^{(0)}{{\cal A}_5^{(1)}}^\ast)$. 
For the Born amplitude we factor out the coupling, we define
\begin{equation}
  {1\over 4} {1\over {\cal N}_c }
  \sum_{\mbox{\scriptsize spin,colour} }|{\cal A}_5^{(0)}|^2
  = \gs^6 a_0.
\end{equation}
The factor $ {1/ {\cal N}_c }$ is due to the average over the incoming
colour. For the channel $\Pg\Pg$, $q\bar q$, $q\Pg$, $\Pg\bar q$ we have
${\cal N}_c = 64,$ 9, 24, 24. 
Note that the coefficient $a_0$ only depends on the chosen phase-space 
point---which implicitly also contains the information about the
top-quark mass.
\def\SubTitle#1{\hline \multicolumn{2}{|c|}{#1}\\ \hline}
\begin{table}[htbp]
  \begin{center}
    \leavevmode
    \begin{tabular}[h]{|c|c|}
      \hline
      &$a_0 [\GeV^{-2}]$ \\
      \SubTitle{$ \Pg\Pg \to \Pt \bar \Pt \Pg $} 
      \Vone& $0.6566843362709776\cdot10^{-3}$\\
      \Vtwo& $0.6566843362709785\cdot10^{-3}$\\
      Madgraph& $0.6566843362709775\cdot10^{-3}$\\
      \SubTitle{$q\bar q\to \Pt\bar\Pt\Pg$}
      \Vone  & $0.5790368001550936\cdot10^{-4}$\\
      \Vtwo  & $0.5790368001550953\cdot10^{-4}$\\
      Madgraph & $0.5790368001550938\cdot10^{-4}$\\
      \SubTitle{$q\Pg\to \Pt\bar\Pt q$}
      \Vone  & $ 0.1607845322071585\cdot10^{-4}$\\
      \Vtwo  & $ 0.1607845322071587\cdot10^{-4}$\\
      Madgraph & $ 0.1607845322071585\cdot10^{-4}$\\
      \SubTitle{$\Pg\bar q\to \Pt\bar\Pt \bar q$}
      \Vone  & $0.2603527972645622\cdot10^{-3}$\\
      \Vtwo  & $0.2603527972645625\cdot10^{-3}$\\
      Madgraph & $0.2603527972645620\cdot10^{-3}$\\
      \hline
   \end{tabular}
    \caption{Colour and spin averaged LO matrix elements squared.}
    \label{tab:bornres}
  \end{center}
\end{table}
The results for $a_0$ are shown in \refta{tab:bornres}.
\Vone and \Vtwo correspond to our implementations. For completeness 
we compare also with Madgraph.
The one-loop
contribution $ {\cal A}^{(1)}$ is renormalised and thus UV finite. 
However, the virtual corrections still contain collinear and soft 
singularities. Similar to what has been done in \citere{Bern:2008ef} 
in a comparison of the virtual corrections to WW+jet production,
we use the decomposition
\begin{equation}
  {1\over 4} {1\over {\cal N}_c }
  \sum_{\mbox{\scriptsize spin,colour} } 
  2\mathrm{Re}({\cal A}_5^{(0)}{{\cal A}_5^{(1)}}^\ast) = \gs^6 a_0\,
  \Gamma(1+\epsilon)\left({4\pi\mu\over \Mt^2 }\right)^\epsilon
  \left(c_{-2}{1\over \epsilon^2} 
    + c_{-1}{1\over \epsilon} 
    + c_0 + {\cal O}(\epsilon) \right).
\end{equation}
The results are shown in \refta{tab:virt}. 
In addition we also give the corresponding results for the 
{\bf I}-operator of the dipole subtraction function as defined in 
\citere{Catani:2002hc}, with the auxiliary parameter $\kappa=2/3$. 
We us the same
decomposition as for the one-loop corrections. The individual
coefficients are shown in \refta{tab:Iop}.
Note that the coefficients $c_i$ with $i=-2,-1,0$ contain one factor of 
$\alpha_{\mathrm{s}}$. 
We use
\begin{equation}
\label{eq:alpha_s}
 \alpha_s(\mu)
 = \frac{4\pi}{\beta_0 L}
 \left( 1 - \frac{\beta_1}{\beta_0^2} \frac{\ln L}{L} \right),
 \;\;\;
 L=\ln(\mu^2/\Lambda^2),
 \;\;\;
 \beta_0 = 11 - \frac{2}{3} N_f,
 \;\;\;
 \beta_1 = 102 - \frac{38}{3} N_f.
\end{equation}
with $N_f=5$ and $\Lambda=226\MeV$, leading to
\begin{equation}
\label{eq:alpha_s_num}
\alpha_{\mathrm{s}}(\Mt) = 0.1075205492734706.
\end{equation}
\def\SubTitle#1{\multicolumn{4}{|c|}{#1}}
\begin{table}[htbp]
  \begin{center}
    \leavevmode
    \begin{tabular}[h]{|c|c|c|c|}
      \hline
      & $c_{-2}$ & $c_{-1}$ & $c_{0}$ \\
      \hline
      \SubTitle{$\Pg\Pg \to \Pt \bar \Pt \Pg $}
      \\
      \hline
     \Vone & $-0.1540118420981379$ & 0.0731096895036588 & 0.5295183452346090  \\
     \Vtwo & $-0.1540118421074573$ & 0.0731096894943437 & 0.5295183452413002  \\
     \hline
      \SubTitle{$q\bar q\to \Pt\bar\Pt\Pg$}\\
     \hline
     \Vone & $-0.0969704191047176$ & $-0.0126983208241891$ & 0.2435672439083931\\
     \Vtwo & $-0.0969704191046950$ & $-0.0126983208241662$ & 0.2435672439081981\\
     \hline
      \SubTitle{$q\Pg\to \Pt\bar\Pt q$}\\
     \hline
     \Vone & $-0.0969704191047088$ & $-0.0056430956994203$ & 0.4003849386477017\\
     \Vtwo & $-0.0969704191046951$ & $-0.0056430956994064$ & 0.4003849386472126\\
     \hline
      \SubTitle{$\Pg\bar q\to \Pt\bar\Pt \bar q$ }\\
     \hline
     \Vone & $-0.0969704191046802$ & 0.0833362739128030 & 0.5384721403213878\\
     \Vtwo & $-0.0969704191046950$ & 0.0833362739127883 & 0.5384721403213897\\
     \hline
    \end{tabular}
    \caption{Coefficients for colour and spin averaged virtual corrections.}
    \label{tab:virt}
  \end{center}
\end{table}
\begin{table}[htbp]
  \begin{center}
    \leavevmode
    \begin{tabular}[h]{|c|c|c|c|}
      \hline
      & $c_{-2}$ & $c_{-1}$ & $c_{0}$ \\
      \hline
      \SubTitle{$\Pg\Pg \to \Pt \bar \Pt \Pg $} \\
      \hline
     \Vone & 0.1540118421074569 &$-0.0731096894943435$&$-0.5280576886301999$ \\
     \Vtwo & 0.1540118421074573 &$-0.0731096894943437$&$-0.5280576886302015$ \\
      \hline
     \SubTitle{$q\bar q\to \Pt\bar\Pt\Pg$} \\
      \hline
     \Vone & 0.0969704191046952 &0.0126983208241661&$-0.3992776407671517$ \\
     \Vtwo & 0.0969704191046950 &0.0126983208241662&$-0.3992776407671513$ \\
      \hline
     \SubTitle{$q\Pg\to \Pt\bar\Pt q$}\\
      \hline
     \Vone & 0.0969704191046950 &0.0056430956994063&$-0.4069645466913195$ \\
     \Vtwo & 0.0969704191046951 &0.0056430956994064&$-0.4069645466913194$ \\
      \hline
      \SubTitle{$\Pg\bar q\to \Pt\bar\Pt \bar q$}\\
      \hline
     \Vone &0.0969704191046950&$-0.0833362739127882$&$-0.3392937280293060$ \\
     \Vtwo &0.0969704191046950&$-0.0833362739127883$&$-0.3392937280293059$\\
      \hline
    \end{tabular}
    \caption{Coefficients for colour and spin averaged results for the {\bf I}-operator. }
    \label{tab:Iop}
  \end{center}
\end{table}
For the LO amplitudes we find an agreement of at least 14 
digits---pretty close to what one can get using 64bit double
precision with 53bits for the mantissa.%
\footnote{Note that some of the code was run on a x86-64
architecture, where usually floating point arithmetic is typically done in the
SMD unit of the processor and thus restricted to 64 bit also for
intermediate results. (On x86 architectures arithmetic is usually done
in the FPU which usually works with extended precision for intermediate
results (80 bit). The precision is reduced when results are stored
back to memory.).}
For the one-loop corrections we find at least an agreement of 10
digits for the finite terms. For the $q\bar q$, $q\Pg$, and $\Pg\bar q$
channels, which are numerically less involved, we find an agreement of up to 14
digits. The finite terms from the {\bf I}-operator agree even better.
The coefficients of all divergences typically agree to 10 digits,
or better in the cases with external quarks.
We note, however, that we do not cancel the IR divergences numerically. 
We also observe a cancellation between the finite part of the 
{\bf I}-operator and the corresponding virtual contributions. For 
the  $\Pg\Pg \to \Pt \bar \Pt \Pg $ channel almost three digits are
cancelled in the combination.

\section{Benchmark numbers for the subtraction terms}
In this section we give also results for the
subtraction term in the dipole formalism
for one phase-space point. We find this useful to
facilitate the comparison of upcoming calculations. Recently
some effort has been invested to automatize this part of the
calculation 
\cite{Gleisberg:2007md,Seymour:2008mu,Hasegawa:2008ae,Frederix:2008hu}. 
The results presented here may provide an interesting benchmark point
for these attempts.
The set of momenta for
$ab\to \Pt\bar\Pt c d$ 
is chosen as
\begin{eqnarray}
p_a &=& \scriptstyle (2100,-0,-0,2100),
\nn\\
p_b &=& \scriptstyle (2800,-0,-0,-2800),
\nn\\
p_{\Pt} &=& \scriptstyle
(1581.118367308447,1254.462316247655,-766.9360998604944,-554.7905976902205),
\nn\\
p_{\bar\Pt} &=& \scriptstyle
(1460.449317799282,-975.9731477430979,-466.5314749495881,965.6402060944737),
\nn\\
p_c &=& \scriptstyle
(545.4084744819,218.7220720302516,472.0439121434804,-163.7241712507502),
\nn\\
p_d &=& \scriptstyle
(1313.023840410371,-497.2112405348086,761.423662666602,-947.1254371535031),
\label{eq:PSpointreal}
\end{eqnarray}
with all components given in GeV. 
The top-quark mass is set to $\Mt=174\GeV$.
We give numbers on the spin- and colour-averaged squared real emission amplitude 
$|{\cal A}^{(0)}_6|^2$ as well as  for the sum of the subtraction terms.
We define the numbers $b_0$ by
\begin{equation}
  {1\over 4} {1\over {\cal N}_c } {1\over {\cal S}}
  \sum_{\mbox{\scriptsize spin,colour} }|{\cal A}^{(0)}_6|^2
  = b_0.
\end{equation}
The factor $1/4$ accounts for the average over the spins of the initial partons,
the factor $ {1/ {\cal N}_c }$ is due to the average over the colour of the incoming
partons. ${\cal N}_c$ contains a factor 8 for every incoming gluon and a factor 3
for every incoming quark or antiquark.
${\cal S}$ is the symmetry factor 
accounting for identical particles in the final state.
The numbers for the dipole subtraction terms are defined analogously:
\begin{equation}
{1\over 4} {1\over {\cal S}} 
\sum\limits_{\mathrm{pairs}\; i,j} \; \sum\limits_{k \neq i,j} {\cal D}_{ij,k} = d_0.
\end{equation}
Note that the factor for the average over the colour of the incoming
partons is included in the definition of ${\cal D}_{ij,k}$.
The numbers $b_0$ and $d_0$ contain the strong coupling constant. As numerical value for
$\alpha_{\mathrm{s}}$ we 
use again \refeq{eq:alpha_s} with $N_f=5$ and $\Lambda=226\MeV$,
i.e.\ the value given in \refeq{eq:alpha_s_num}.
\def\SubTitle#1{\hline \multicolumn{3}{|c|}{#1}\\ \hline}
\begin{table}[htbp]
  \begin{center}
    \leavevmode
    \begin{tabular}[h]{|c|c|c|}
      \hline
      &$b_0 [\GeV^{-4}]$ &$d_0 [\GeV^{-4}]$ \\
      \SubTitle{$ \Pg(p_a)\Pg(p_b) \to \Pt(p_{\Pt}) \bar \Pt(p_{\bar\Pt}) \Pg(p_c) \Pg(p_d) $} 
      \Vone  & $ 3.12815868347843 \cdot 10^{-9}$ & $ 4.1037601540955 \cdot 10^{-9}$ \\
      \Vtwo  & $ 3.12815868347842 \cdot 10^{-9}$ & $ 4.1037601540962 \cdot 10^{-9}$ \\
      \SubTitle{$ q(p_a)\bar q(p_b) \to \Pt(p_{\Pt}) \bar \Pt(p_{\bar\Pt}) \Pg(p_c) \Pg(p_d) $} 
      \Vone  & $ 4.48308845446477 \cdot 10^{-10}$ & $ 4.90476067759639 \cdot 10^{-10}$ \\
      \Vtwo  & $ 4.48308845446475 \cdot 10^{-10}$ & $ 4.90476067759631 \cdot 10^{-10}$ \\
      \SubTitle{$ q(p_a)\Pg(p_b) \to \Pt(p_{\Pt}) \bar \Pt(p_{\bar\Pt}) \Pg(p_c) q(p_d) $} 
      \Vone  & $ 1.10256509258713 \cdot 10^{-10}$ & $ 1.919073353538 \cdot 10^{-10}$ \\
      \Vtwo  & $ 1.10256509258713 \cdot 10^{-10}$ & $ 1.919073353539 \cdot 10^{-10}$ \\
      \SubTitle{$ \bar q(p_a)\Pg(p_b) \to \Pt(p_{\Pt}) \bar \Pt(p_{\bar\Pt}) \bar q(p_c) \Pg(p_d) $} 
      \Vone  & $ 1.384600673183816 \cdot 10^{-10}$ & $ 3.3382231835799 \cdot 10^{-10}$ \\
      \Vtwo  & $ 1.384600673183812 \cdot 10^{-10}$ & $ 3.3382231835798 \cdot 10^{-10}$ \\
      \SubTitle{$ \Pg(p_a)\Pg(p_b) \to \Pt(p_{\Pt}) \bar \Pt(p_{\bar\Pt}) \bar q(p_c) q(p_d) $} 
      \Vone  & $ 2.42841040229558 \cdot 10^{-10}$ & $ 4.271065781530 \cdot 10^{-10}$ \\
      \Vtwo  & $ 2.42841040229557 \cdot 10^{-10}$ & $ 4.271065781532 \cdot 10^{-10}$ \\
      \hline
   \end{tabular}
    \caption{Colour and spin averaged real emission matrix element squared and dipole subtraction terms
             related to the processes $0 \to \Pt \bar \Pt \Pg \Pg \Pg \Pg$ and $0 \to \Pt \bar \Pt q \bar q \Pg \Pg$.}
    \label{tab:real1}
  \end{center}
\end{table}
\begin{table}[htbp]
  \begin{center}
    \leavevmode
    \begin{tabular}[h]{|c|c|c|}
      \hline
      &$b_0 [\GeV^{-4}]$ &$d_0 [\GeV^{-4}]$ \\
      \SubTitle{$ q(p_a) q'(p_b) \to \Pt(p_{\Pt}) \bar \Pt(p_{\bar\Pt}) q'(p_c) q(p_d) $} 
      \Vone  & $ 4.44137855516180 \cdot 10^{-12}$ & $ 1.6381811832266 \cdot 10^{-11}$ \\
      \Vtwo  & $ 4.44137855516180 \cdot 10^{-12}$ & $ 1.6381811832275 \cdot 10^{-11}$ \\
      \SubTitle{$ \bar q(p_a) \bar q'(p_b) \to \Pt(p_{\Pt}) \bar \Pt(p_{\bar\Pt}) \bar q(p_c) \bar q'(p_d) $} 
      \Vone  & $ 1.733763330485899 \cdot 10^{-11}$ & $ 1.06832579841007 \cdot 10^{-10}$ \\
      \Vtwo  & $ 1.733763330485899 \cdot 10^{-11}$ & $ 1.06832579841000 \cdot 10^{-10}$ \\
      \SubTitle{$ q(p_a)\bar q'(p_b) \to \Pt(p_{\Pt}) \bar \Pt(p_{\bar\Pt}) \bar q'(p_c) q(p_d) $} 
      \Vone  & $ 4.796260245409952 \cdot 10^{-12}$ & $ 1.8776008214791 \cdot 10^{-11}$ \\
      \Vtwo  & $ 4.796260245409957 \cdot 10^{-12}$ & $ 1.8776008214799 \cdot 10^{-11}$ \\
      \SubTitle{$ q(p_a)\bar q(p_b) \to \Pt(p_{\Pt}) \bar \Pt(p_{\bar\Pt}) \bar q'(p_c) q'(p_d) $} 
      \Vone  & $ 6.13924303047741 \cdot 10^{-11}$ & $ 6.990891152615 \cdot 10^{-11}$ \\
      \Vtwo  & $ 6.13924303047739 \cdot 10^{-11}$ & $ 6.990891152614 \cdot 10^{-11}$ \\
      \SubTitle{$ q(p_a)q(p_b) \to \Pt(p_{\Pt}) \bar \Pt(p_{\bar\Pt}) q(p_c) q(p_d) $} 
      \Vone  & $ 1.371477814148721 \cdot 10^{-11}$ & $ 4.1848434402744 \cdot 10^{-11}$ \\
      \Vtwo  & $ 1.371477814148719 \cdot 10^{-11}$ & $ 4.1848434402750 \cdot 10^{-11}$ \\
      \SubTitle{$ \bar q(p_a)\bar q(p_b) \to \Pt(p_{\Pt}) \bar \Pt(p_{\bar\Pt}) \bar q(p_c) \bar q(p_d) $} 
      \Vone  & $ 1.411042000289490 \cdot 10^{-11}$ & $ 6.3674516988854 \cdot 10^{-11}$ \\
      \Vtwo  & $ 1.411042000289488 \cdot 10^{-11}$ & $ 6.3674516988857 \cdot 10^{-11}$ \\
      \SubTitle{$ q(p_a)\bar q(p_b) \to \Pt(p_{\Pt}) \bar \Pt(p_{\bar\Pt}) \bar q(p_c) q(p_d) $} 
      \Vone  & $ 2.054843839960259 \cdot 10^{-11}$ & $ 3.6253236096328 \cdot 10^{-11}$ \\
      \Vtwo  & $ 2.054843839960252 \cdot 10^{-11}$ & $ 3.6253236096334 \cdot 10^{-11}$ \\
      \hline
   \end{tabular}
    \caption{Colour and spin averaged real emission matrix element squared and dipole subtraction terms
             related to the processes $0 \to \Pt \bar \Pt q \bar q q' \bar q'$ and $0 \to \Pt \bar \Pt q \bar q q \bar q$.}
    \label{tab:real2}
  \end{center}
\end{table}
The results for $b_0$ and $d_0$ are shown in \reftas{tab:real1} and 
\ref{tab:real2}.
The two implementations agree at least to 14 digits for the matrix elements squared and at least to 12 digits
for the sum of the subtraction terms.

\section{Tables for histograms}
In this appendix we give the tables for the differential distributions.
For each distribution, we list the NLO predictions for
the scale choice $\mu = \Mt/2$, $\mu = \Mt$ and $\mu = 2\Mt$.
In all tables we have set $\mu=\mu_{\mathrm{ren}}=\mu_{\mathrm{fact}}$.
The errors result from the Monte Carlo integration. The bin is
specified by its central value. The bin width---which we chose
constant for the entire histogram---is obtained from the distance of
two neighboring bin positions. 
Note that we use the same definition for the cross section as
described in \refse{sect:results}. In particular, we demand a minimum
$p_{\rT}$ for the additional jet. For the Tevatron $20\GeV$ is used while
for the LHC $50\GeV$ is used.
\def\arraystretch{0.9}
\def\FBoverGeV{[\frac{\mbox{\scriptsize fb}}{\mbox{\scriptsize GeV}}]}
\def\dsigma#1{{\frac{d\sigma}{#1}}}
\begin{table}[htbp]
  \begin{center}
    \begin{tabular}{|r|c|c|c|}
      \hline
      & \multicolumn{3}{|c|}{$\dsigma{dp_{\rT,\mathrm{jet}}} \FBoverGeV$} \\
      $p_{\rT,\mathrm{jet}}$ [GeV] & $\mu=\Mt/2$ & $\mu=\Mt$ & $\mu=2\Mt$ \\
      \hline
      $12.5$  &  $16.83 \pm 0.07$    &  $15.70 \pm 0.04$     &   $13.4 \pm 0.2$    \\
      $37.5$  &  $38.47 \pm 0.04$    &  $35.08 \pm 0.03$     &   $28.83 \pm 0.02$    \\
      $62.5$  &  $12.67 \pm 0.02$    &  $11.45 \pm 0.01$     &   $9.340 \pm 0.009$    \\
      $87.5$  &  $5.29 \pm 0.01$     &  $4.805 \pm 0.008$    &   $3.92 \pm 0.005$    \\
      $112.5$ &  $2.465 \pm 0.007$   &  $2.277 \pm 0.006$    &   $1.864 \pm 0.003$    \\
      $137.5$ &  $1.216 \pm 0.005$   &  $1.146 \pm 0.003$    &   $0.943 \pm 0.002$    \\
      $162.5$ &  $0.629 \pm 0.004$   &  $0.604 \pm 0.002$    &   $0.496 \pm 0.001$    \\
      $187.5$ &  $0.326 \pm 0.003$   &  $0.324 \pm 0.002$    &   $0.2659 \pm 0.0007$   \\
      $212.5$ &  $0.173 \pm 0.002$   &  $0.174 \pm 0.001$    &   $0.1460 \pm 0.0005$   \\
      $237.5$ &  $0.093 \pm 0.001$   &  $0.0945 \pm 0.0008$  &   $0.0796 \pm 0.0004$   \\
      $262.5$ &  $0.047 \pm 0.001$   &  $0.0522 \pm 0.0006$  &   $0.0445 \pm 0.0003$   \\
      $287.5$ &  $0.0252 \pm 0.0007$ &  $0.0285 \pm 0.0003$  &   $0.0238 \pm 0.0002$   \\
      $312.5$ &  $0.0113 \pm 0.0006$ &  $0.0151 \pm 0.0003$  &   $0.0129 \pm 0.0001$   \\
      \hline
    \end{tabular}
    \caption{\label{table:pt_Tevatron_jet}
      The transverse momentum distribution of the hard jet at the Tevatron.
      }
  \end{center}
\end{table}
\begin{table}[htbp]
\begin{center}
  \begin{tabular}{|r|c|c|c|}
    \hline
    & \multicolumn{3}{|c|}{$\dsigma{dp_{\rT,\Pt\bar\Pt}}\FBoverGeV $} \\
    $p_{\rT,\Pt\bar\Pt} [\mbox{GeV}]$ & $\mu=\Mt/2$ & $\mu=\Mt$ & $\mu=2\Mt$ \\
    \hline
    $12.5$  &     $16.43 \pm 0.07$    & $15.49 \pm 0.04$    & $13.2 \pm 0.2$      \\
    $37.5$  &     $37.25 \pm 0.05$    & $34.43 \pm 0.03$    & $28.46 \pm 0.02$    \\
    $62.5$  &     $13.56 \pm 0.02$    & $11.91 \pm 0.01$    & $9.61 \pm 0.01$     \\
    $87.5$  &     $5.66 \pm 0.01$     & $5.036 \pm 0.007$   & $4.053 \pm 0.004$   \\
    $112.5$  &    $2.648 \pm 0.007$   & $2.373 \pm 0.005$   & $1.919 \pm 0.003$   \\
    $137.5$  &    $1.309 \pm 0.006$   & $1.194 \pm 0.003$   & $0.967 \pm 0.002$   \\
    $162.5$  &    $0.669 \pm 0.004$   & $0.623 \pm 0.002$   & $0.509 \pm 0.001$   \\
    $187.5$  &    $0.345 \pm 0.002$   & $0.332 \pm 0.001$   & $0.2720 \pm 0.0007$ \\
    $212.5$  &    $0.180 \pm 0.002$   & $0.1791 \pm 0.0008$ & $0.1482 \pm 0.0005$ \\
    $237.5$  &    $0.096 \pm 0.001$   & $0.0953 \pm 0.0006$ & $0.0797 \pm 0.0004$ \\
    $262.5$  &    $0.0466 \pm 0.0009$ & $0.0525 \pm 0.0005$ & $0.0443 \pm 0.0002$ \\
    $287.5$  &    $0.0243 \pm 0.0007$ & $0.0280 \pm 0.0003$ & $0.0239 \pm 0.0002$ \\
    $312.5$  &    $0.0113 \pm 0.0005$ & $0.0144 \pm 0.0003$ & $0.0126 \pm 0.0001$ \\
    \hline
  \end{tabular}
  \caption{\label{table:pt_Tevatron_ttbar}
    The transverse momentum distribution of the total $\Pt\bar\Pt$ system at the Tevatron.
    }
\end{center}
\end{table}
\begin{table}[htbp]
\begin{center}
\begin{tabular}{|r|c|c|c|}
\hline
 & \multicolumn{3}{|c|}{$\dsigma{dp_{\rT,\Pt}} \FBoverGeV$} \\
 $p_{\rT,\Pt} [\mbox{GeV}]$ & $\mu=\Mt/2$ & $\mu=\Mt$ & $\mu=2\Mt$ \\
\hline
   $12.5$  &     $4.12 \pm 0.02$   & $3.58 \pm 0.01$   & $2.910 \pm 0.007$ \\
   $37.5$  &     $10.96 \pm 0.03$  & $9.62 \pm 0.02$   & $7.78 \pm 0.02$   \\
   $62.5$  &     $14.31 \pm 0.03$  & $12.68 \pm 0.02$  & $10.35 \pm 0.02$  \\
   $87.5$  &     $14.11 \pm 0.03$  & $12.63 \pm 0.02$  & $10.58 \pm 0.02$  \\
   $112.5$  &    $11.67 \pm 0.03$  & $10.66 \pm 0.02$  & $8.72 \pm 0.01$   \\
   $137.5$  &    $8.53 \pm 0.02$   & $7.94 \pm 0.01$   & $6.59 \pm 0.01$   \\
   $162.5$  &    $5.75 \pm 0.02$   & $5.48 \pm 0.01$   & $4.570 \pm 0.008$ \\
   $187.5$  &    $3.59 \pm 0.03$   & $3.59 \pm 0.01$   & $3.019 \pm 0.008$ \\
   $212.5$  &    $2.23 \pm 0.02$   & $2.24 \pm 0.01$   & $1.923 \pm 0.006$ \\
   $237.5$  &    $1.30 \pm 0.01$   & $1.391 \pm 0.008$ & $1.189 \pm 0.004$ \\
   $262.5$  &    $0.76 \pm 0.01$   & $0.837 \pm 0.006$ & $0.721 \pm 0.004$ \\
   $287.5$  &    $0.43 \pm 0.01$   & $0.488 \pm 0.004$ & $0.428 \pm 0.003$ \\
   $312.5$  &    $0.237 \pm 0.009$ & $0.286 \pm 0.004$ & $0.254 \pm 0.003$ \\
   $337.5$  &    $0.139 \pm 0.006$ & $0.167 \pm 0.003$ & $0.146 \pm 0.002$ \\
   $362.5$  &    $0.067 \pm 0.005$ & $0.088 \pm 0.003$ & $0.082 \pm 0.002$ \\
   $387.5$  &    $0.032 \pm 0.005$ & $0.051 \pm 0.002$ & $0.047 \pm 0.001$ \\
\hline
\end{tabular}
\caption{\label{table:pt_Tevatron_top}
The transverse momentum distribution of the top-quark at the Tevatron.
}
\end{center}
\end{table}
\begin{table}[htbp]
\begin{center}
\begin{tabular}{|r|c|c|c|}
\hline
 & \multicolumn{3}{|c|}{$\dsigma{d\eta_{\Pt}} [\mbox{fb}]$} \\
 $\eta_{\Pt}$ & $\mu=\Mt/2$ & $\mu=\Mt$ & $\mu=2\Mt$ \\
\hline 
 $-3.8$ & $6.6 \pm 0.2$   & $6.3 \pm 0.1$   & $5.2 \pm 0.1$ \\
 $-3.4$ & $16.0 \pm 0.3$  & $13.7 \pm 0.2$  & $11.2 \pm 0.1$ \\
 $-3.0$ & $33.5 \pm 0.4$  & $30.1 \pm 0.3$  & $24.1 \pm 0.2$ \\
 $-2.6$ & $70.1 \pm 0.8$  & $62.5 \pm 0.3$  & $50.9 \pm 0.2$ \\
 $-2.2$ & $136.4 \pm 0.8$ & $123.3 \pm 0.5$ & $100.3 \pm 0.3$ \\
 $-1.8$ & $237 \pm 1$     & $219.6 \pm 0.9$ & $188 \pm 9$ \\
 $-1.4$ & $363 \pm 1$     & $338.1 \pm 0.8$ & $279.6 \pm 0.5$ \\
 $-1.0$ & $483 \pm 1$     & $451.8 \pm 0.8$ & $376.2 \pm 0.6$ \\
 $-0.6$ & $546 \pm 2$     & $519 \pm 1$     & $437 \pm 2$ \\
 $-0.2$ & $573 \pm 2$     & $543.8 \pm 0.9$ & $454.5 \pm 0.6$ \\
 $0.2$  & $573 \pm 2$     & $534.3 \pm 0.8$ & $446.2 \pm 0.6$ \\
 $0.6$  & $542 \pm 1$     & $496.6 \pm 0.9$ & $409.1 \pm 0.6$ \\
 $1.0$  & $467 \pm 1$     & $420.7 \pm 0.8$ & $341.4 \pm 0.6$ \\
 $1.4$  & $354 \pm 1$     & $310 \pm 1$     & $248.6 \pm 0.5$ \\
 $1.8$  & $229 \pm 1$     & $197.2 \pm 0.5$ & $165 \pm 9$ \\
 $2.2$  & $130.5 \pm 0.7$ & $109.4 \pm 0.4$ & $86.4 \pm 0.3$ \\
 $2.6$  & $65.9 \pm 0.5$  & $55.3 \pm 0.3$  & $43.0 \pm 0.2$ \\
 $3.0$  & $32.4 \pm 0.4$  & $26.6 \pm 0.2$  & $20.9 \pm 0.2$ \\
 $3.4$  & $15.2 \pm 0.3$  & $12.4 \pm 0.1$  & $9.7 \pm 0.1$ \\
 $3.8$  & $6.7 \pm 0.2$   & $5.7 \pm 0.1$   & $4.42 \pm 0.08$ \\
\hline
\end{tabular}
\caption{\label{table:eta_Tevatron_top}
The pseudo-rapidity distribution of the top-quark at the Tevatron.
}
\end{center}
\end{table}
\begin{table}[htbp]
\begin{center}
\begin{tabular}{|r|c|c|c|}
\hline
 & \multicolumn{3}{|c|}{$\dsigma{dy_{\Pt}} [\mbox{fb}]$} \\
 $y_{\Pt}$ & $\mu=\Mt/2$ & $\mu=\Mt$ & $\mu=2\Mt$ \\
\hline 
 $-1.8$ & $9.8 \pm 0.3$   & $8.5 \pm 0.1$   & $6.9 \pm 0.1$ \\
 $-1.4$ & $103.5 \pm 0.8$ & $96.3 \pm 0.4$  & $78.5 \pm 0.3$ \\
 $-1.0$ & $388 \pm 1$     & $368.4 \pm 0.8$ & $314 \pm 9$ \\
 $-0.6$ & $826 \pm 2$     & $777 \pm 2$     & $645.0 \pm 0.8$ \\
 $-0.2$ & $1144 \pm 2$    & $1063 \pm 1$    & $888 \pm 2$ \\
 $0.2$  & $1131 \pm 3$    & $1035 \pm 1$    & $856.6 \pm 0.8$ \\
 $0.6$  & $799 \pm 2$     & $716 \pm 1$     & $581.9 \pm 0.8$ \\
 $1.0$  & $379 \pm 1$     & $329.4 \pm 0.9$ & $270 \pm 9$ \\
 $1.4$  & $100.9 \pm 0.6$ & $82.9 \pm 0.3$  & $63.5 \pm 0.2$ \\
 $1.8$  & $10.1 \pm 0.1$  & $7.68 \pm 0.08$ & $5.46 \pm 0.07$ \\
\hline
\end{tabular}
\caption{\label{table:y_Tevatron_top}
The rapidity distribution of the top-quark at the Tevatron.
}
\end{center}
\end{table}
\begin{table}[htbp]
\begin{center}
\begin{tabular}{|r|c|c|c|}
\hline
 & \multicolumn{3}{|c|}{$\dsigma{dy_{\mathrm{jet}}} [\mbox{fb}]$} \\
 $y_{\mathrm{jet}}$ & $\mu=\Mt/2$ & $\mu=\Mt$ & $\mu=2\Mt$ \\
\hline 
 $-3.8$ & $1.5 \pm 0.1$   & $1.26 \pm 0.09$  & $1.0 \pm 0.1$ \\
 $-3.4$ & $10.8 \pm 0.4$  & $9.4 \pm 0.2$    & $7.5 \pm 0.2$ \\
 $-3.0$ & $41.1 \pm 0.7$  & $35.8 \pm 0.4$   & $28.5 \pm 0.3$ \\
 $-2.6$ & $100 \pm 1$     & $88.0 \pm 0.6$   & $70.3 \pm 0.4$ \\
 $-2.2$ & $192 \pm 1$     & $168.2 \pm 0.7$  & $135.0 \pm 0.5$ \\
 $-1.8$ & $295 \pm 1$     & $266 \pm 1$      & $215.5 \pm 0.6$ \\
 $-1.4$ & $397 \pm 2$     & $357 \pm 1$      & $293.3 \pm 0.6$ \\
 $-1.0$ & $456 \pm 2$     & $422 \pm 1$      & $348.4 \pm 0.6$ \\
 $-0.6$ & $481 \pm 2$     & $447 \pm 1$      & $380 \pm 9$ \\
 $-0.2$ & $471 \pm 2$     & $449 \pm 2$      & $374.4 \pm 0.7$ \\
 $0.2$  & $475 \pm 2$     & $448 \pm 2$      & $374.9 \pm 0.7$ \\
 $0.6$  & $477 \pm 2$     & $445 \pm 2$      & $380 \pm 9$ \\
 $1.0$  & $457 \pm 2$     & $421 \pm 1$      & $348.8 \pm 0.5$ \\
 $1.4$  & $396 \pm 2$     & $359 \pm 1$      & $292.8 \pm 0.5$ \\
 $1.8$  & $294 \pm 1$     & $265 \pm 1$      & $214.9 \pm 0.5$ \\
 $2.2$  & $191 \pm 2$     & $167.9 \pm 0.6$  & $135.3 \pm 0.5$ \\
 $2.6$  & $101.1 \pm 0.9$ & $89.0 \pm 0.5$   & $70.1 \pm 0.4$ \\
 $3.0$  & $41.8 \pm 0.6$  & $36.1 \pm 0.4$   & $28.7 \pm 0.3$ \\
 $3.4$  & $11.3 \pm 0.3$  & $9.8 \pm 0.2$    & $7.7 \pm 0.2$ \\
 $3.8$  & $1.5 \pm 0.1$   & $1.16 \pm 0.09$  & $0.92 \pm 0.07$ \\
\hline
\end{tabular}
\caption{\label{table:y_Tevatron_jet}
The rapidity distribution of the hard jet at the Tevatron.
}
\end{center}
\end{table}
%
%
%
%
\begin{table}[htbp]
\begin{center}
\begin{tabular}{|r|c|c|c|}
\hline
 & \multicolumn{3}{|c|}{$\dsigma{dp_{\rT,\mathrm{jet}}} \FBoverGeV$} \\
 $p_{\rT,\mathrm{jet}} [\mbox{GeV}]$ & $\mu=\Mt/2$ & $\mu=\Mt$ & $\mu=2\Mt$ \\
\hline
 $25$  & $0$             & $0$             & $0$              \\
 $75$  & $4201 \pm 11$   & $4045 \pm 5$    & $3553 \pm 4$     \\
 $125$ & $1762 \pm 5$    & $1635 \pm 2$    & $1405 \pm 2$     \\
 $175$ & $863 \pm 3$     & $802 \pm 2$     & $685 \pm 1$      \\
 $225$ & $454 \pm 3$     & $428 \pm 1$     & $368.0 \pm 0.8$  \\
 $275$ & $254 \pm 2$     & $242.7 \pm 0.7$ & $210.9 \pm 0.5$  \\
 $325$ & $144 \pm 1$     & $142.1 \pm 0.5$ & $124.9 \pm 0.4$  \\
 $375$ & $85.9 \pm 0.7$  & $86.8 \pm 0.4$  & $76.2 \pm 0.3$   \\
 $425$ & $51.5 \pm 0.5$  & $53.8 \pm 0.3$  & $47.7 \pm 0.2$   \\
 $475$ & $31.3 \pm 0.6$  & $34.1 \pm 0.2$  & $30.6 \pm 0.1$   \\
 $525$ & $19.1 \pm 0.5$  & $21.8 \pm 0.3$  & $20.0 \pm 0.1$   \\
 $575$ & $12.6 \pm 0.2$  & $14.7 \pm 0.2$  & $13.01 \pm 0.08$ \\
 $625$ & $7.4 \pm 0.2$   & $9.4 \pm 0.1$   & $9.06 \pm 0.06$  \\
 $675$ & $4.7 \pm 0.1$   & $6.40 \pm 0.09$ & $6.07 \pm 0.05$  \\
\hline
\end{tabular}
\caption{\label{table:pt_LHC_jet}
The transverse momentum distribution of the hard jet at the LHC.
}
\end{center}
\end{table}
\begin{table}[htbp]
\begin{center}
\begin{tabular}{|r|c|c|c|}
\hline
 & \multicolumn{3}{|c|}{$\dsigma{dp_{\rT,\Pt\bar\Pt}} \FBoverGeV$} \\
 $p_{\rT,\Pt\bar\Pt} [\mbox{GeV}]$ & $\mu=\Mt/2$ & $\mu=\Mt$ & $\mu=2\Mt$ \\
\hline
 $25$  & $2390 \pm 10$  & $1457 \pm 6$    & $920 \pm 2$     \\
 $75$  & $1512 \pm 11$  & $2418 \pm 5$    & $2524 \pm 4$    \\
 $125$ & $1915 \pm 5$   & $1722 \pm 3$    & $1460 \pm 1$    \\
 $175$ & $938 \pm 3$    & $843 \pm 2$     & $714 \pm 1$     \\
 $225$ & $496 \pm 2$    & $454 \pm 1$     & $383.9 \pm 0.7$ \\
 $275$ & $274 \pm 1$    & $254.4 \pm 0.7$ & $217.9 \pm 0.5$ \\
 $325$ & $157 \pm 1$    & $148.7 \pm 0.5$ & $128.2 \pm 0.4$ \\
 $375$ & $88 \pm 1$     & $89.2 \pm 0.4$  & $77.7 \pm 0.3$  \\
 $425$ & $54.5 \pm 0.8$ & $54.9 \pm 0.3$  & $48.3 \pm 0.2$  \\
 $475$ & $31.6 \pm 0.6$ & $34.1 \pm 0.2$  & $30.7 \pm 0.2$  \\
 $525$ & $18.8 \pm 0.6$ & $21.4 \pm 0.2$  & $19.8 \pm 0.1$  \\
 $575$ & $11.5 \pm 0.2$ & $13.3 \pm 0.2$  & $13.07 \pm 0.09$\\
 $625$ & $6.6 \pm 0.2$  & $9.2 \pm 0.2$   & $8.50 \pm 0.08$ \\
 $675$ & $3.7 \pm 0.1$  & $5.91 \pm 0.07$ & $5.72 \pm 0.06$ \\
\hline
\end{tabular}
\caption{\label{table:pt_LHC_ttbar}
The transverse momentum distribution of the total $\Pt\bar\Pt$ system at the LHC.
}
\end{center}
\end{table}
\begin{table}[htbp]
\begin{center}
\begin{tabular}{|r|c|c|c|}
\hline
 & \multicolumn{3}{|c|}{$\dsigma{dp_{\rT,\Pt}} \FBoverGeV$} \\
 $p_{\rT,\Pt} [\mbox{GeV}]$ & $\mu=\Mt/2$ & $\mu=\Mt$ & $\mu=2\Mt$ \\
\hline
 $25$  & $995 \pm 3$    & $881 \pm 2$      & $740 \pm 2$     \\
 $75$  & $2152 \pm 5$   & $1930 \pm 4$     & $1639 \pm 2$    \\
 $125$ & $2029 \pm 6$   & $1858 \pm 3$     & $1597 \pm 3$    \\
 $175$ & $1338 \pm 11$  & $1276 \pm 2$     & $1115 \pm 2$    \\
 $225$ & $723 \pm 7$    & $744 \pm 2$      & $662 \pm 2$     \\
 $275$ & $368 \pm 2$    & $400 \pm 1$      & $365.8 \pm 0.9$ \\
 $325$ & $178 \pm 2$    & $209.3 \pm 0.8$  & $196.4 \pm 0.5$ \\
 $375$ & $83 \pm 1$     & $110.5 \pm 0.6$  & $107.5 \pm 0.6$ \\
 $425$ & $37.9 \pm 0.7$ & $58.4 \pm 0.4$   & $60.1 \pm 0.5$  \\
 $475$ & $14.9 \pm 0.6$ & $31.8 \pm 0.3$   & $33.8 \pm 0.2$  \\
 $525$ & $4.5 \pm 0.4$  & $17.5 \pm 0.2$   & $19.6 \pm 0.1$  \\
 $575$ & $0.1 \pm 0.3$  & $9.6 \pm 0.2$    & $11.4 \pm 0.2$  \\
 $625$ & --             & $5.4 \pm 0.1$    & $6.6 \pm 0.1$   \\
 $675$ & --             & $3.0 \pm 0.1$    & $4.21 \pm 0.07$ \\
\hline
\end{tabular}
\caption{\label{table:pt_LHC_top}
The transverse momentum distribution of the top-quark at the LHC.
}
\end{center}
\end{table}
\begin{table}[htbp]
\begin{center}
\begin{tabular}{|r|c|c|c|}
\hline
 & \multicolumn{3}{|c|}{$\dsigma{d\eta_{\Pt}} [\mbox{pb}]$} \\
 $\eta_{\Pt}$ & $\mu=\Mt/2$ & $\mu=\Mt$ & $\mu=2\Mt$ \\
\hline
 $-5.7$ & $0.27 \pm 0.04$ & $0.21 \pm 0.01$ & $0.184 \pm 0.008$ \\
 $-5.1$ & $0.85 \pm 0.04$ & $0.71 \pm 0.03$ & $0.58 \pm 0.01$ \\
 $-4.5$ & $2.58 \pm 0.07$ & $2.27 \pm 0.03$ & $1.92 \pm 0.04$ \\
 $-3.9$ & $7.5 \pm 0.1$   & $6.55 \pm 0.05$ & $5.35 \pm 0.05$ \\
 $-3.3$ & $18.3 \pm 0.1$  & $16.3 \pm 0.1$  & $13.54 \pm 0.07$ \\
 $-2.7$ & $36.4 \pm 0.2$  & $32.8 \pm 0.1$  & $27.86 \pm 0.07$ \\
 $-2.1$ & $56.1 \pm 0.2$  & $52.2 \pm 0.1$  & $45.05 \pm 0.09$ \\
 $-1.5$ & $69.6 \pm 0.2$  & $66.7 \pm 0.2$  & $58.3 \pm 0.1$ \\
 $-0.9$ & $70.4 \pm 0.5$  & $69.7 \pm 0.1$  & $61.64 \pm 0.09$ \\
 $-0.3$ & $66.6 \pm 0.2$  & $66.7 \pm 0.1$  & $59.28 \pm 0.09$ \\
 $0.3$  & $67.0 \pm 0.2$  & $66.5 \pm 0.1$  & $59.39 \pm 0.09$ \\
 $0.9$  & $71.4 \pm 0.3$  & $69.5 \pm 0.3$  & $61.57 \pm 0.09$ \\
 $1.5$  & $69.6 \pm 0.2$  & $66.6 \pm 0.3$  & $58.1 \pm 0.1$ \\
 $2.1$  & $56.3 \pm 0.3$  & $52.8 \pm 0.3$  & $45.10 \pm 0.08$ \\
 $2.7$  & $36.0 \pm 0.2$  & $32.6 \pm 0.2$  & $27.80 \pm 0.08$ \\
 $3.3$  & $18.3 \pm 0.1$  & $16.3 \pm 0.1$  & $13.43 \pm 0.08$ \\
 $3.9$  & $7.8 \pm 0.2$   & $6.57 \pm 0.05$ & $5.43 \pm 0.04$ \\
 $4.5$  & $2.6 \pm 0.1$   & $2.28 \pm 0.04$ & $1.82 \pm 0.03$ \\
 $5.1$  & $0.83 \pm 0.04$ & $0.70 \pm 0.02$ & $0.61 \pm 0.01$ \\
 $5.7$  & $0.25 \pm 0.02$ & $0.20 \pm 0.01$ & $0.171 \pm 0.009$ \\
\hline
\end{tabular}
\caption{\label{table:eta_LHC_top}
The pseudo-rapidity distribution of the top-quark at the LHC.
}
\end{center}
\end{table}
\begin{table}[htbp]
\begin{center}
\begin{tabular}{|r|c|c|c|}
\hline
 & \multicolumn{3}{|c|}{$\dsigma{dy_{\Pt}} [\mbox{pb}]$} \\
 $y_{\Pt}$ & $\mu=\Mt/2$ & $\mu=\Mt$ & $\mu=2\Mt$ \\
\hline 
 $-3.9$ & $0.093 \pm 0.008$ & $0.075 \pm 0.006$ & $0.053 \pm 0.004$ \\
 $-3.3$ & $2.39 \pm 0.06$   & $1.97 \pm 0.04$   & $1.55 \pm 0.02$ \\
 $-2.7$ & $12.9 \pm 0.1$    & $11.55 \pm 0.09$  & $9.52 \pm 0.05$ \\
 $-2.1$ & $35.7 \pm 0.2$    & $33.3 \pm 0.1$    & $28.35 \pm 0.07$ \\
 $-1.5$ & $66.7 \pm 0.2$    & $63.3 \pm 0.2$    & $55.1 \pm 0.1$ \\
 $-0.9$ & $96.5 \pm 0.3$    & $92.9 \pm 0.2$    & $81.5 \pm 0.1$ \\
 $-0.3$ & $114.1 \pm 0.6$   & $111.1 \pm 0.2$   & $97.7 \pm 0.1$ \\
 $0.3$  & $115.6 \pm 0.3$   & $110.9 \pm 0.2$   & $97.7 \pm 0.1$ \\
 $0.9$  & $96.5 \pm 0.2$    & $92.9 \pm 0.2$    & $81.4 \pm 0.2$ \\
 $1.5$  & $67.0 \pm 0.3$    & $63.5 \pm 0.2$    & $55.0 \pm 0.1$ \\
 $2.1$  & $35.6 \pm 0.2$    & $33.1 \pm 0.1$    & $28.32 \pm 0.07$ \\
 $2.7$  & $12.8 \pm 0.1$    & $11.45 \pm 0.07$  & $9.50 \pm 0.06$ \\
 $3.3$  & $2.32 \pm 0.08$   & $1.95 \pm 0.03$   & $1.56 \pm 0.02$ \\
 $3.9$  & $0.095 \pm 0.006$ & $0.071 \pm 0.005$ & $0.050 \pm 0.003$ \\
\hline
\end{tabular}
\caption{\label{table:y_LHC_top}
The rapidity distribution of the top-quark at the LHC.
}
\end{center}
\end{table}
\begin{table}[htbp]
\begin{center}
\begin{tabular}{|r|c|c|c|}
\hline
 & \multicolumn{3}{|c|}{$\dsigma{dy_{\mathrm{jet}}} [\mbox{pb}]$} \\
 $y_{\mathrm{jet}}$ & $\mu=\Mt/2$ & $\mu=\Mt$ & $\mu=2\Mt$ \\
\hline 
 $-5.1$ & $0.05 \pm 0.04$ & $0.18 \pm 0.02$ & $0.12 \pm 0.01$ \\
 $-4.5$ & $2.2 \pm 0.1$   & $2.09 \pm 0.07$ & $1.88 \pm 0.04$ \\
 $-3.9$ & $9.8 \pm 0.2$   & $9.8 \pm 0.1$   & $8.34 \pm 0.07$ \\
 $-3.3$ & $23.0 \pm 0.2$  & $22.2 \pm 0.1$  & $19.38 \pm 0.08$ \\
 $-2.7$ & $39.1 \pm 0.3$  & $37.1 \pm 0.1$  & $32.21 \pm 0.08$ \\
 $-2.1$ & $53.3 \pm 0.5$  & $50.5 \pm 0.1$  & $44.01 \pm 0.09$ \\
 $-1.5$ & $62.6 \pm 0.5$  & $59.9 \pm 0.2$  & $52.2 \pm 0.1$ \\
 $-0.9$ & $68.3 \pm 0.6$  & $65.1 \pm 0.2$  & $56.8 \pm 0.1$ \\
 $-0.3$ & $71.1 \pm 0.3$  & $67.2 \pm 0.1$  & $58.8 \pm 0.1$ \\
 $0.3$  & $70.7 \pm 0.3$  & $67.0 \pm 0.3$  & $58.8 \pm 0.1$ \\
 $0.9$  & $68.1 \pm 0.3$  & $65.4 \pm 0.3$  & $56.8 \pm 0.2$ \\
 $1.5$  & $63.0 \pm 0.3$  & $60.1 \pm 0.1$  & $52.1 \pm 0.1$ \\
 $2.1$  & $53.6 \pm 0.3$  & $50.5 \pm 0.1$  & $44.0 \pm 0.1$ \\
 $2.7$  & $39.1 \pm 0.3$  & $36.9 \pm 0.1$  & $32.18 \pm 0.09$ \\
 $3.3$  & $22.8 \pm 0.2$  & $22.3 \pm 0.1$  & $19.3 \pm 0.1$ \\
 $3.9$  & $9.9 \pm 0.1$   & $10.2 \pm 0.4$  & $8.34 \pm 0.06$ \\
 $4.5$  & $2.09 \pm 0.09$ & $2.2 \pm 0.1$   & $1.88 \pm 0.06$ \\
 $5.1$  & $0.12 \pm 0.02$ & $0.13 \pm 0.02$ & $0.12 \pm 0.01$ \\
\hline
\end{tabular}
\caption{\label{table:y_LHC_jet}
The rapidity distribution of the hard jet at the LHC.
}
\end{center}
\end{table}

\bibliography{ppttj}
\bibliographystyle{h-physrev3}

\end{document}